\documentclass{SCGE}
\begin{document}

\begin{picture}(0,0){\rm
\put(0,-39){\makebox[160truemm][l]{\bf {\sanhao\raisebox{2pt}{.}}
Article {\sanhao\raisebox{1.5pt}{.}}}}}
\put(0,-52){\jiuwuhao {\textcolor[rgb]{0.5,0.5,0.5}{\sf 
}}}
\end{picture}

\def\bm{\boldsymbol}

\def\dl{\displaystyle}
\def\du{\end{document}}
\def\pi{{\uppi}}

\Year{2013} %
\Month{??} %
\Vol{??} %
\No{??} %
\BeginPage{1} %
\EndPage{??} %
\AuthorMark{{\rm S.~M. Liang et al.:} Discovery of eight lensing clusters of galaxies}
\DOI{??} 

\title{Discovery of eight lensing clusters of galaxies}

\author[1,2]{S.~M. Liang}{}%
\author[1]{Z.~L. Wen*}{}
\author[1]{J.~L. Han}{}
\author[1]{Y.~Y. Jiang}{}

\address[{\rm1}]{National Astronomical Observatories, Chinese Academy
  of Sciences, 20A Datun Road, Chaoyang District, Beijing 100012,
  China}
\address[{\rm2}]{The University of Chinese Academy of Sciences,
  Beijing 100049, China} 

\maketitle \vspace{-3.5mm}{\footnotesize\begin{center} 
Received August 1st, 2013; accepted ?? ??, 2013
\end{center}}\vspace*{-5mm}

\begin{center}
\rule{16.5cm}{0.4pt}
\parbox{16.5cm}
{\begin{abstract} 
Clusters of galaxies have a huge mass which can act as gravitational
lenses. Galaxies behind clusters can be distorted to form arcs in
images by the lenses. Herein a search was done for giant lensed arcs
by galaxy clusters using the SDSS data. By visually inspecting SDSS
images of newly identified clusters in the SDSS DR8 and Stripe 82
data, we discover 8 strong lensing clusters together with additional 3
probable and 6 possible cases. The lensed arcs show bluer colors than
the member galaxies of clusters. The masses and optical
luminosities of galaxy clusters interior to the arcs are calculated, 
and the mass-to-light ratios are found to be in the range of a few tens of
$M_{\odot}/L_{\odot}$, consistent with the distribution of previously
known lensing clusters.
\end{abstract}}
\end{center}\vspace*{-0.6cm}

\begin{center}
\parbox{16.5cm}
{\bf\jiuhao key words: Galaxy clusters, Gravitational lenses and luminous arcs,
Photography and photometry}%
\end{center}

\begin{center}
{}%
\vspace*{-1mm}
\Cit{S.~M. Liang et al., Discovery of eight lensing clusters of galaxies.
 Sci China-Phys Mech Astron, 2013, 55: 1--??, doi: ?? } 
\end{center}

\wuhao\vspace*{1.5mm}

\begin{multicols}{2}

\renewcommand{\baselinestretch}{1.08} \baselineskip 12.2pt\parindent=10.8pt

\renewcommand{\thefootnote}

\sec{1\quad Introduction}

Galaxy clusters are the largest gravitational bound systems in the
universe, with a mass of about
10$^{14}$~$M_{\odot}$--10$^{15}$~$M_{\odot}$.  About $\sim$3\% of
such a huge mass comes from galaxies visible in optical,
$\sim$15\% of the mass exists in the form of diffuse hot
intracluster gas detected in X-ray, and about $\sim$82\% of mass is
contributed from invisible matter, probably dark matter [1].

Galaxy clusters with such a mass have a huge gravity that can deflect
lights of background sources, so that they act as lenses. When a
background galaxy is located very close to the light-of-sight of a
lens in the sky, it can be magnified to be a brighter object but
distorted to form giant arcs or multiple images. The properties of
lensed arcs, such as locations and shapes, depend on the mass
distribution of the gravitational lens and are independent of the form
of matter (luminous or dark, gas or stars). Therefore, the lensing
effect is a unique tool to measure the total mass of a galaxy
cluster. Such measured mass of galaxy clusters can be used to
calibrate other observable mass proxies in optical and X-ray [2],
which can be easily applied to a complete sample of galaxy 
%
%
\noindent\rule{2.5cm}{0.4pt}\\[0.1mm]{\qihao *Corresponding author (email:
zhonglue@nao.cas.cn) 
} \\
\no
clusters to
calculate the mass function and then constraint cosmology parameters
[3].

Previously, many efforts have been made to search for giant lensed
arcs in various survey data based on samples of massive clusters, for
example, Einstein Observatory Extended Medium Sensitivity Survey [4],
Red-Sequence Cluster Survey data [5], Hubble WFPC2 data [6],
Canada-France-Hawaii Telescope Legacy Survey data [7], Sloan Digital
Sky Survey (SDSS) data [8--11], the Wisconsin-Indiana-Yale NOAO 3.5m
telescope and the University of Hawaii 88 inch telescope data
[12]. Up to now, about 318 strong lensing clusters of galaxies
have been discovered.

\begin{table*}
%
\small
\caption{Parameters of newly discovered lensing clusters of galaxies}
\setlength{\tabcolsep}{1.5mm}
\label{tab1}
\centering
\begin{tabular*}{0.99\textwidth}{lccccrcccccr}
\hline 
\hline
Cluster Position & $z$ &Richness&$\theta _e$&$r$ &$(g-r)_{\rm arc}$&$(g-r)_{\rm lens}$& 
Sky & Lensing & ${\rm M_e}$&${\rm M_e/L_e}$\\
&&&($''$)&(mag)&(mag)&(mag)&region &Probability&$({10^{13}\rm M_{\odot}})$&$({\rm M_{\odot}/L_{\odot}})$\\
\hline
 
011107.2+134644  &0.595&25.00&8.8&22.14$\pm$0.14&0.05$\pm$0.24&1.04$\pm$0.14&DR8      &almost certain&2.52& 79.4\\
012237.5+011123  &0.550&29.18&6.5&22.46$\pm$0.28&0.89$\pm$0.53&1.94$\pm$0.19&Stripe 82&almost certain&1.25& 35.2\\
022510.8$-$004707&0.380&20.89&9.6&22.92$\pm$0.10&0.89$\pm$1.63&1.52$\pm$0.09&Stripe 82&almost certain&1.79&132.8\\
092049.9+452158  &0.661&51.98&4.1&20.59$\pm$0.05&0.73$\pm$0.09&2.03$\pm$0.30&DR8      &almost certain&0.63& 48.3\\
120535.4+411044  &0.662&94.30&3.4&20.65$\pm$0.06&0.39$\pm$0.08&3.20$\pm$0.76&DR8      &almost certain&0.43&  7.5\\ 
150924.7+390140  &0.593&20.59&3.0&22.03$\pm$0.14&0.08$\pm$0.17&1.83$\pm$0.20&DR8      &almost certain&0.29&  6.3\\
152559.9+084639  &0.602&50.23& 3.4&22.16$\pm$0.12&$-0.40\pm$0.13&1.24$\pm$0.24&DR8    &almost certain&0.48& 16.1\\
162320.3+215535  &0.482&23.85& 6.0&20.89$\pm$0.11& 0.20$\pm$0.14&1.53$\pm$0.06&DR8    &almost certain&0.91& 36.8\\[2mm]

025932.5+001354  &0.209& 58.00&14.8&19.81$\pm$0.21& 0.18$\pm$0.32&1.38$\pm$0.02&Stripe 82 & probable&2.32&55.5\\ 
033304.7$-$065122&0.635&151.62& 8.9&22.29$\pm$0.22&$-0.01\pm$0.26&3.17$\pm$0.88&DR8           &probable&2.81&25.4\\
231354.5$-$010449&0.546& 53,75& 7.9&22.27$\pm$0.23& 0.31$\pm$0.35&1.87$\pm$0.31&Stripe 82     &probable&1.83&45.3\\[2mm]

021342.9$-$000359&0.693& 19.42& 4.0&23.86$\pm$0.17&$-0.13\pm$0.20&1.32$\pm$0.11&Stripe 82     &possible&0.64&69.4\\
023830.7+004855  &0.406& 16.36& 2.5&23.21$\pm$0.49&$-0.42\pm$0.54&1.55$\pm$0.08&Stripe 82     &possible&0.13&17.9\\
032812.7+003309  &0.471& 11.37& 4.6&21.85$\pm$0.18& 1.65$\pm$0.67&1.73$\pm$0.20&Stripe 82     &possible&0.52&34.9\\
140217.1+392820  &0.602& 85.45& 4.1&21.71$\pm$0.10& 0.50$\pm$0.15&1.86$\pm$0.15&DR8           &possible&0.56&17.0\\
224140.9$-$005750&0.496& 30.63& 2.6&21.90$\pm$0.25&$-0.11\pm$0.32&2.25$\pm$0.24&Stripe 82     &possible&0.18&10.0\\
234709.1$-$000457&0.263& 13.62& 4.9&22.11$\pm$0.27& 0.24$\pm$0.32&1.02$\pm$0.03&Stripe 82     &possible&0.32&45.7\\
\hline
\end{tabular*}
\end{table*}

\begin{figure*}
  \begin{tabular}{ccc}
\includegraphics[width = 2.1in]{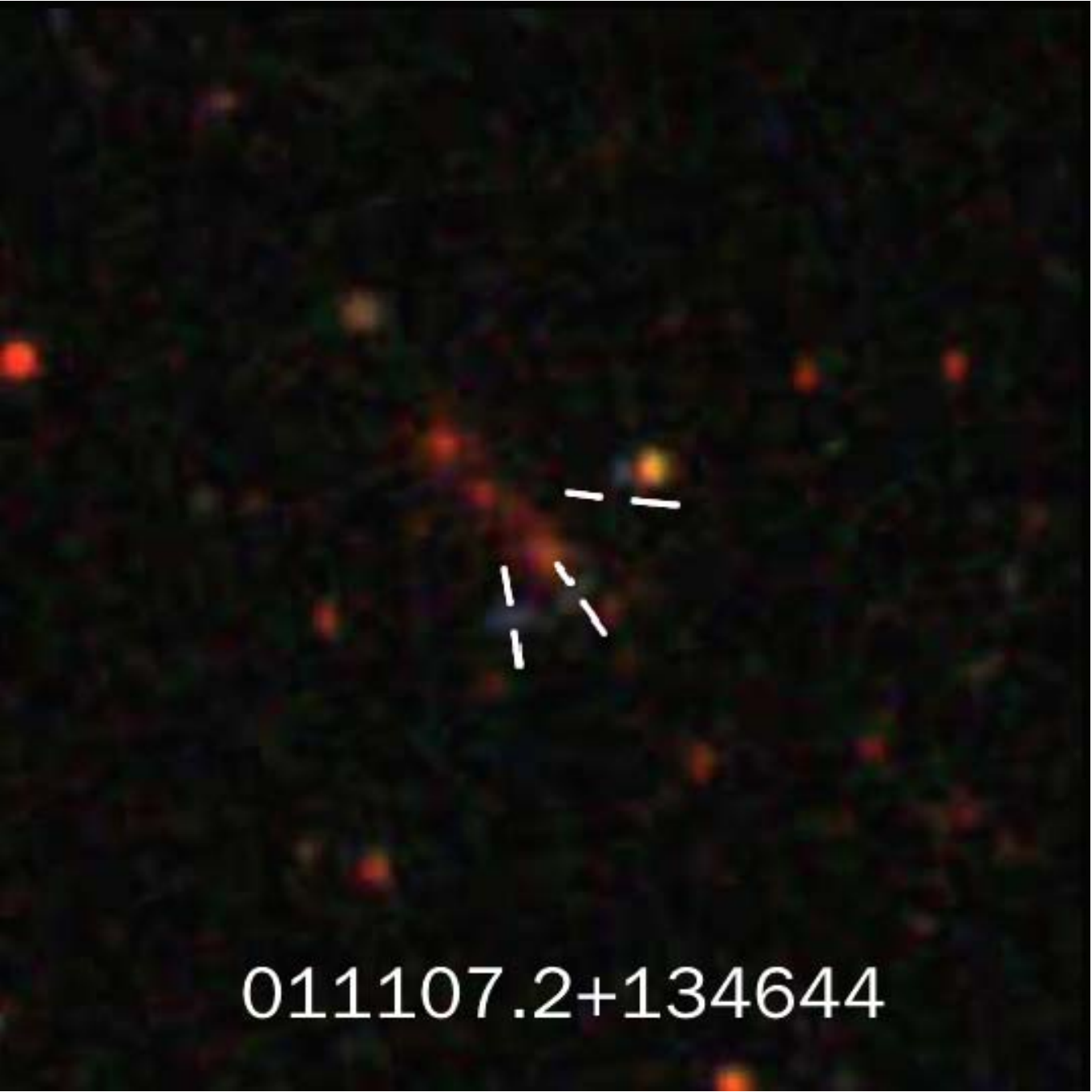}&
\includegraphics[width = 2.1in]{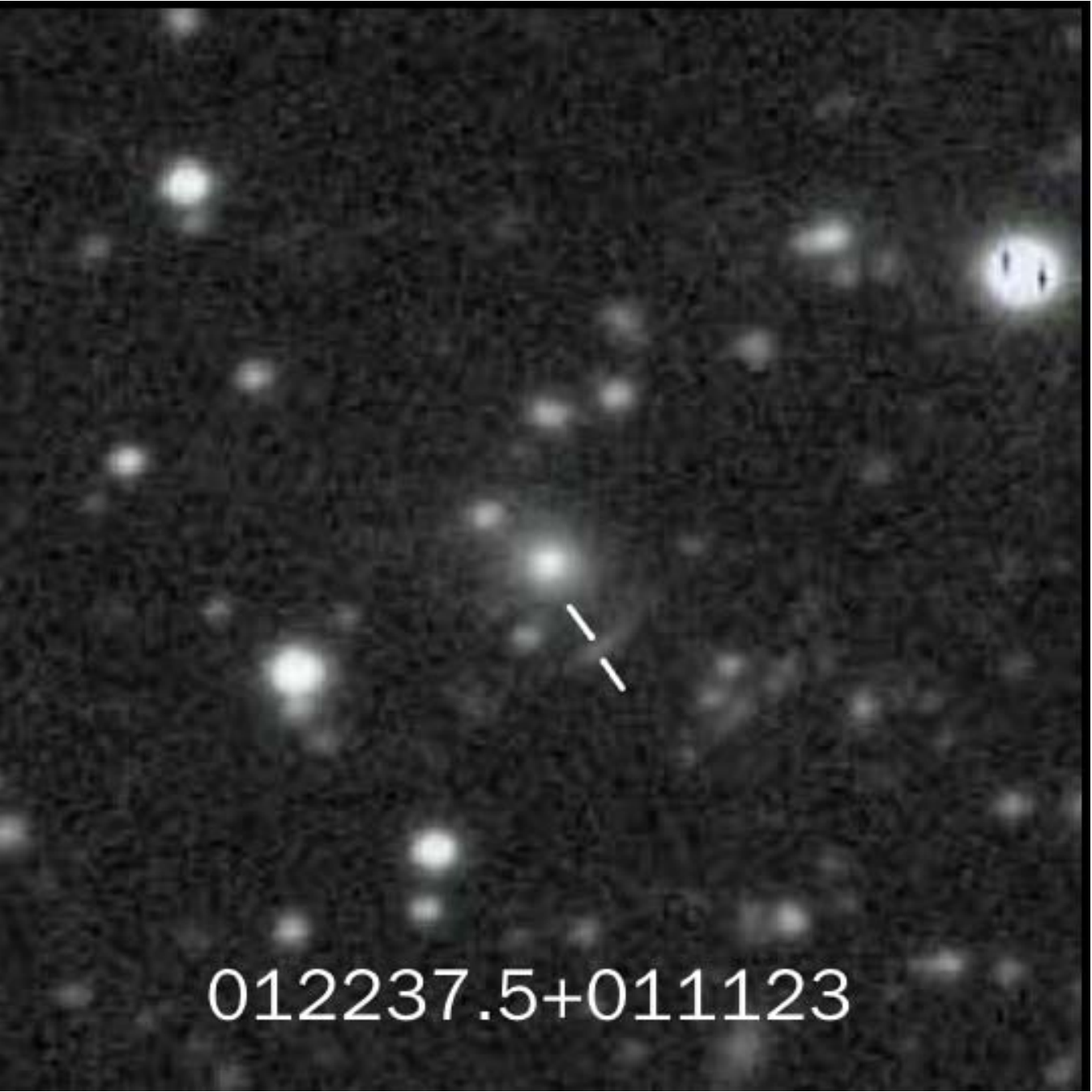}&
\includegraphics[width = 2.1in]{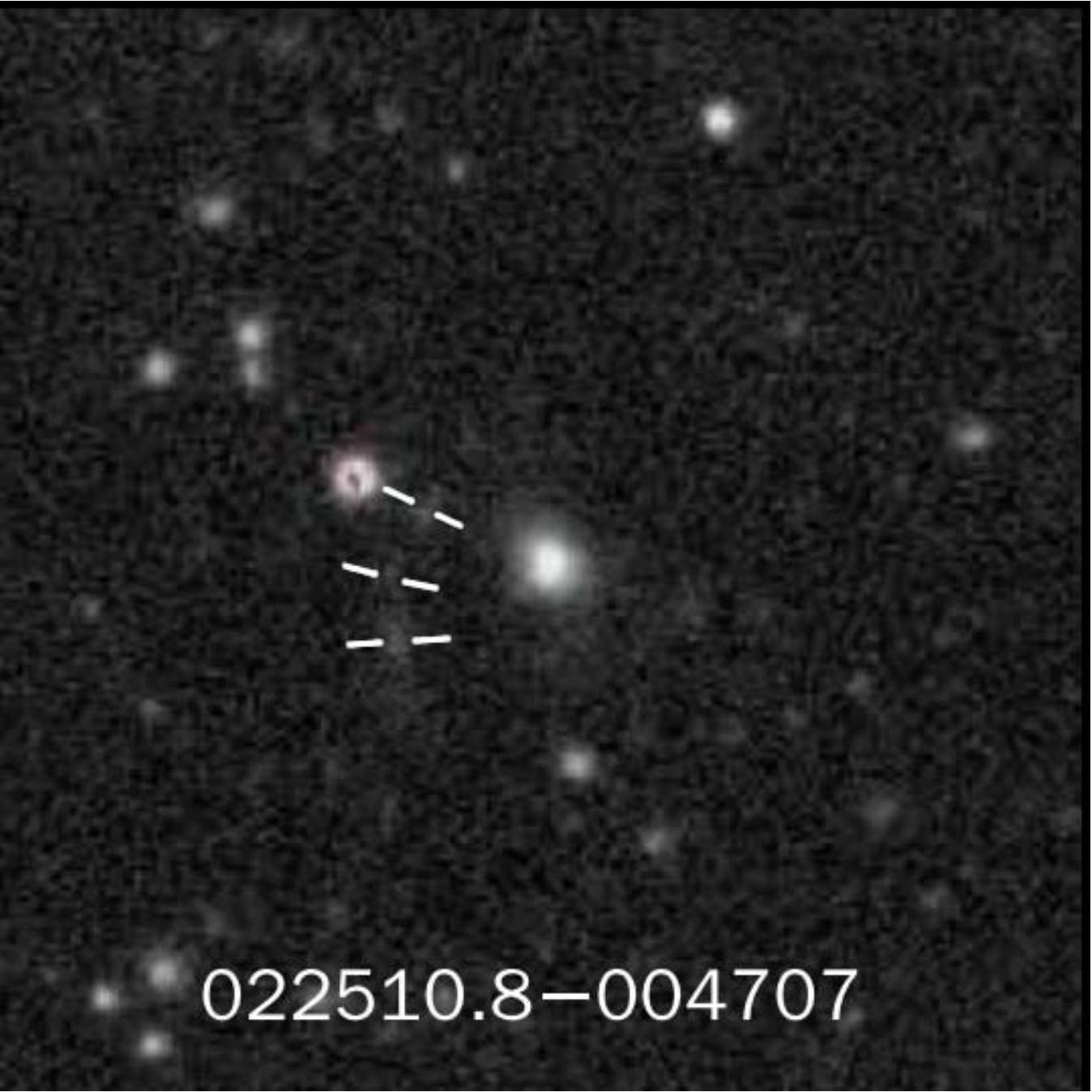}\\
\includegraphics[width = 2.1in]{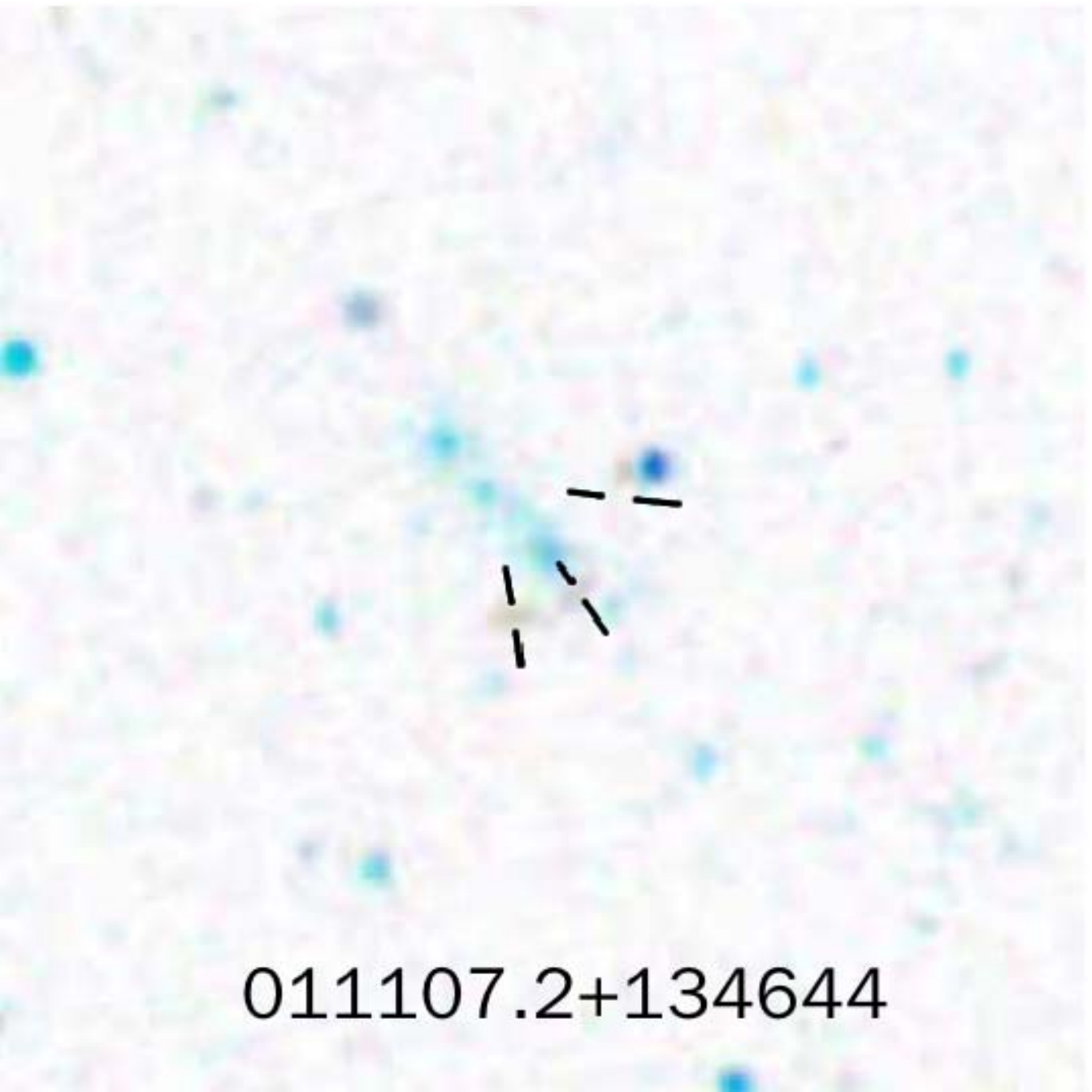}&
\includegraphics[width = 2.1in]{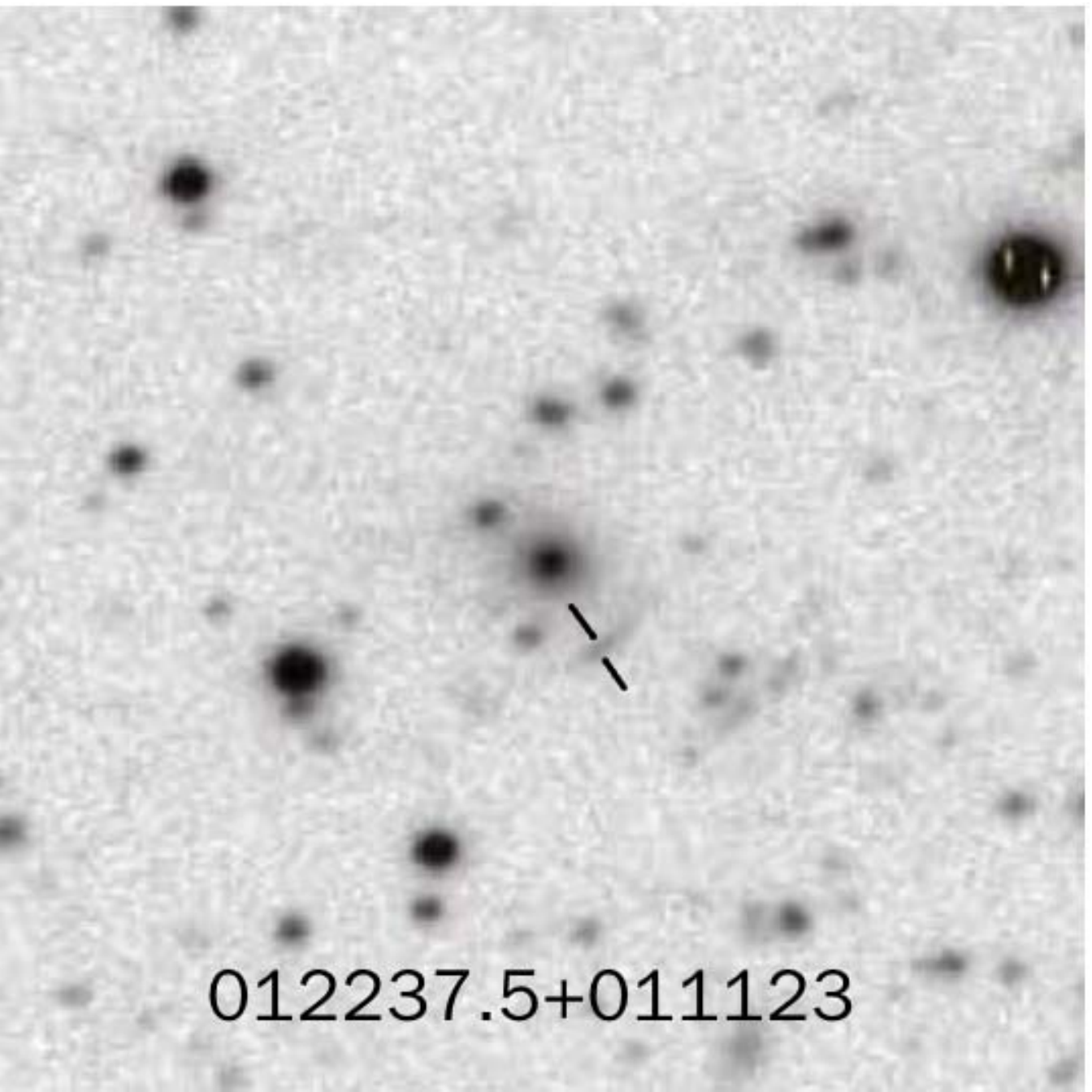}&
\includegraphics[width = 2.1in]{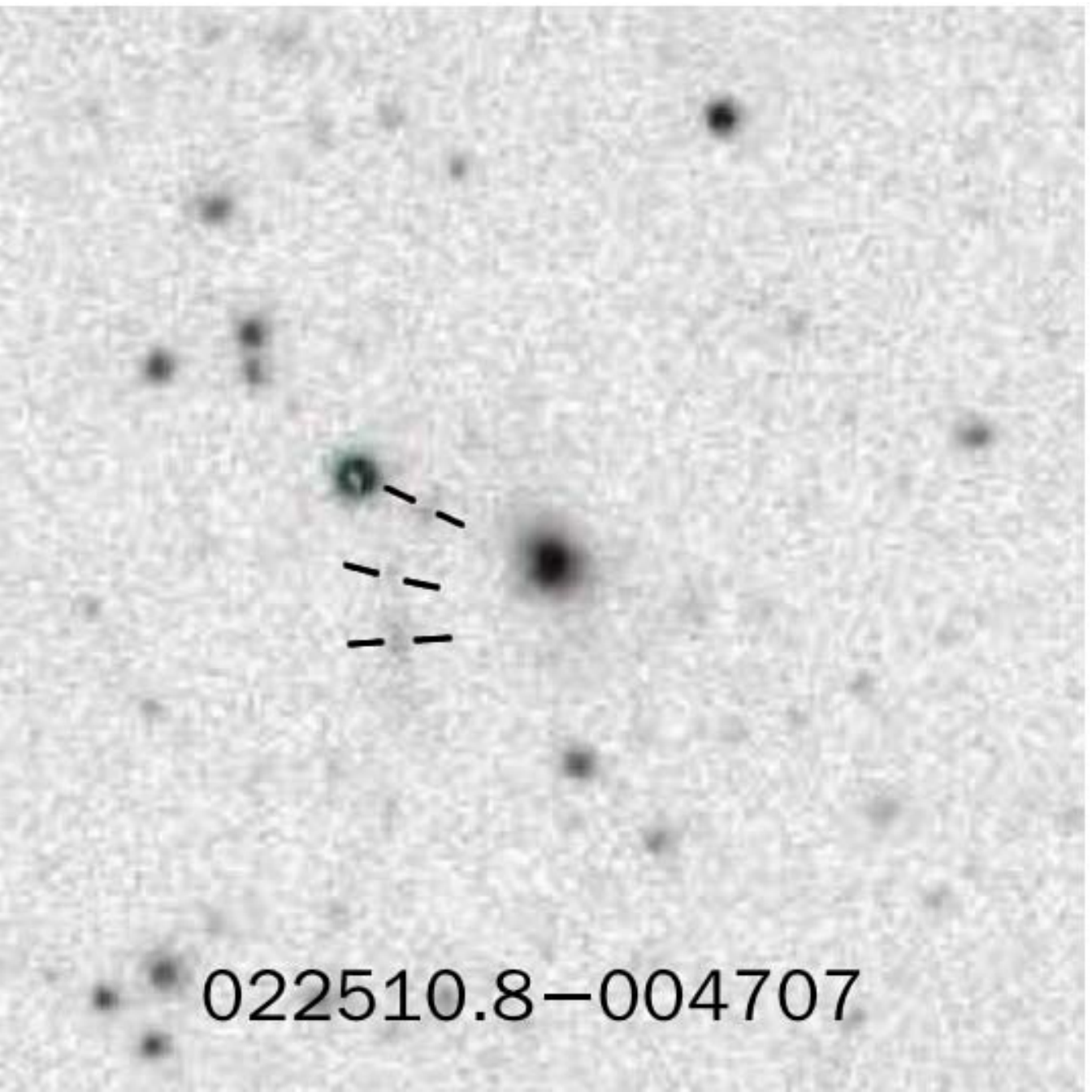}\\
\includegraphics[width = 2.1in]{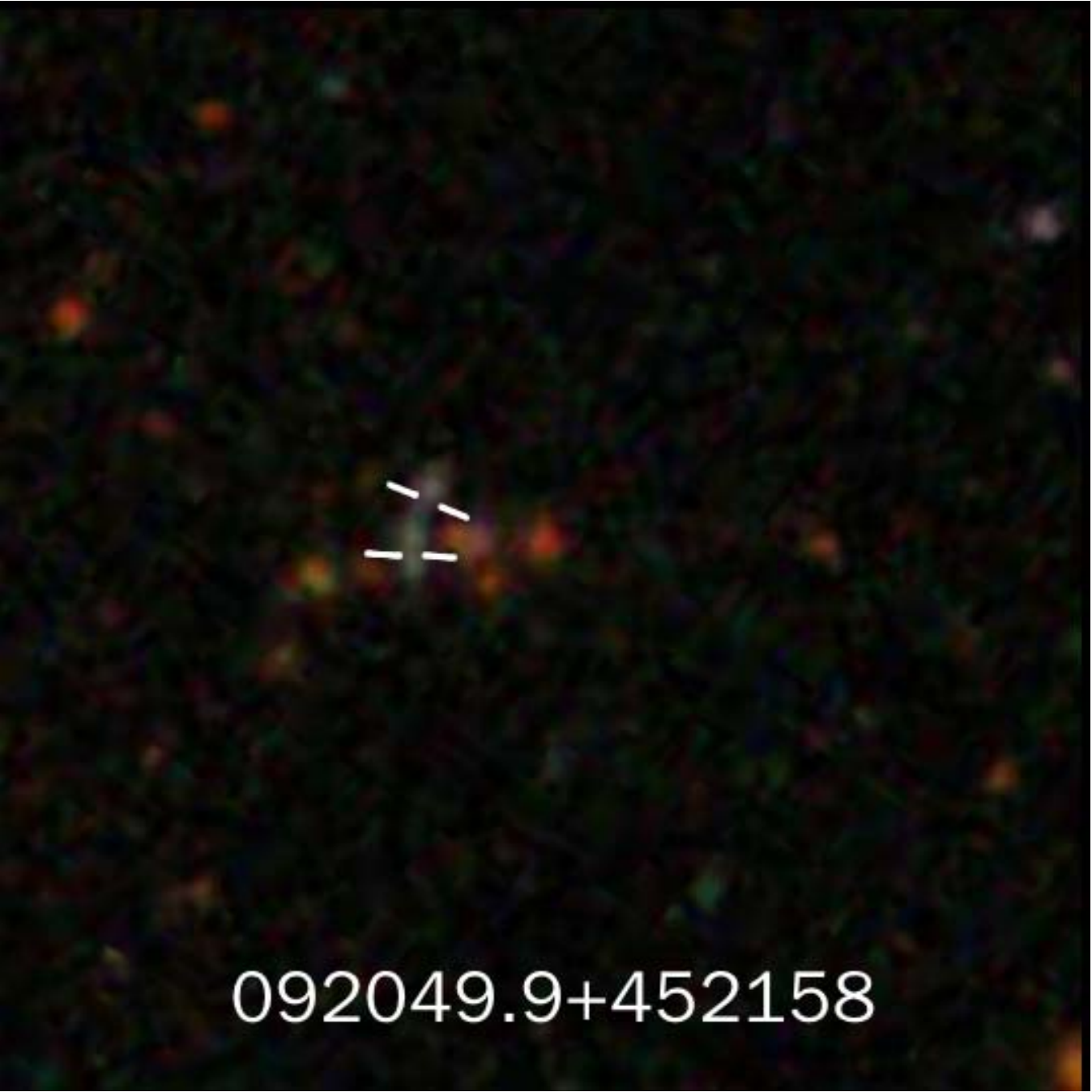}&
\includegraphics[width = 2.1in]{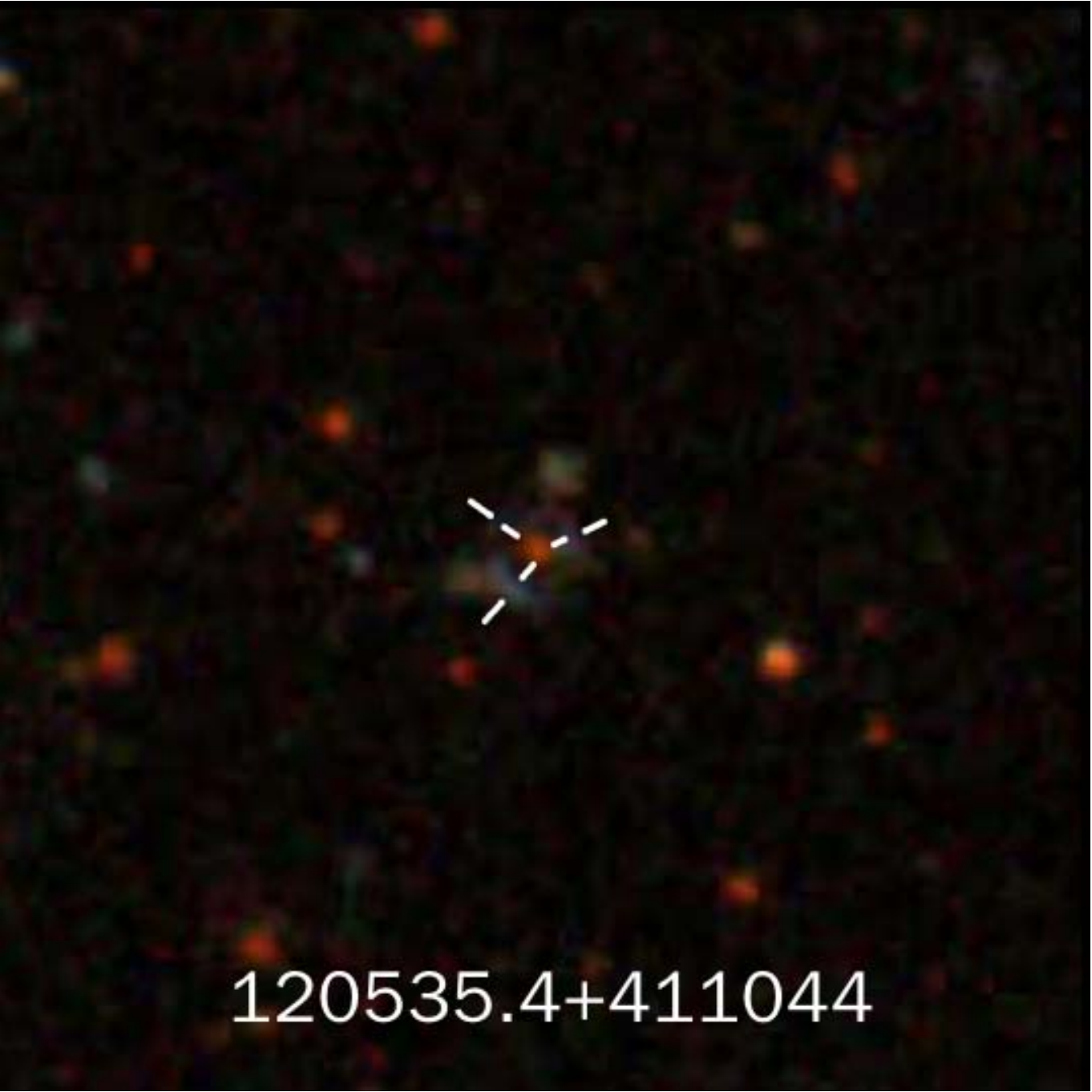}&
\includegraphics[width = 2.1in]{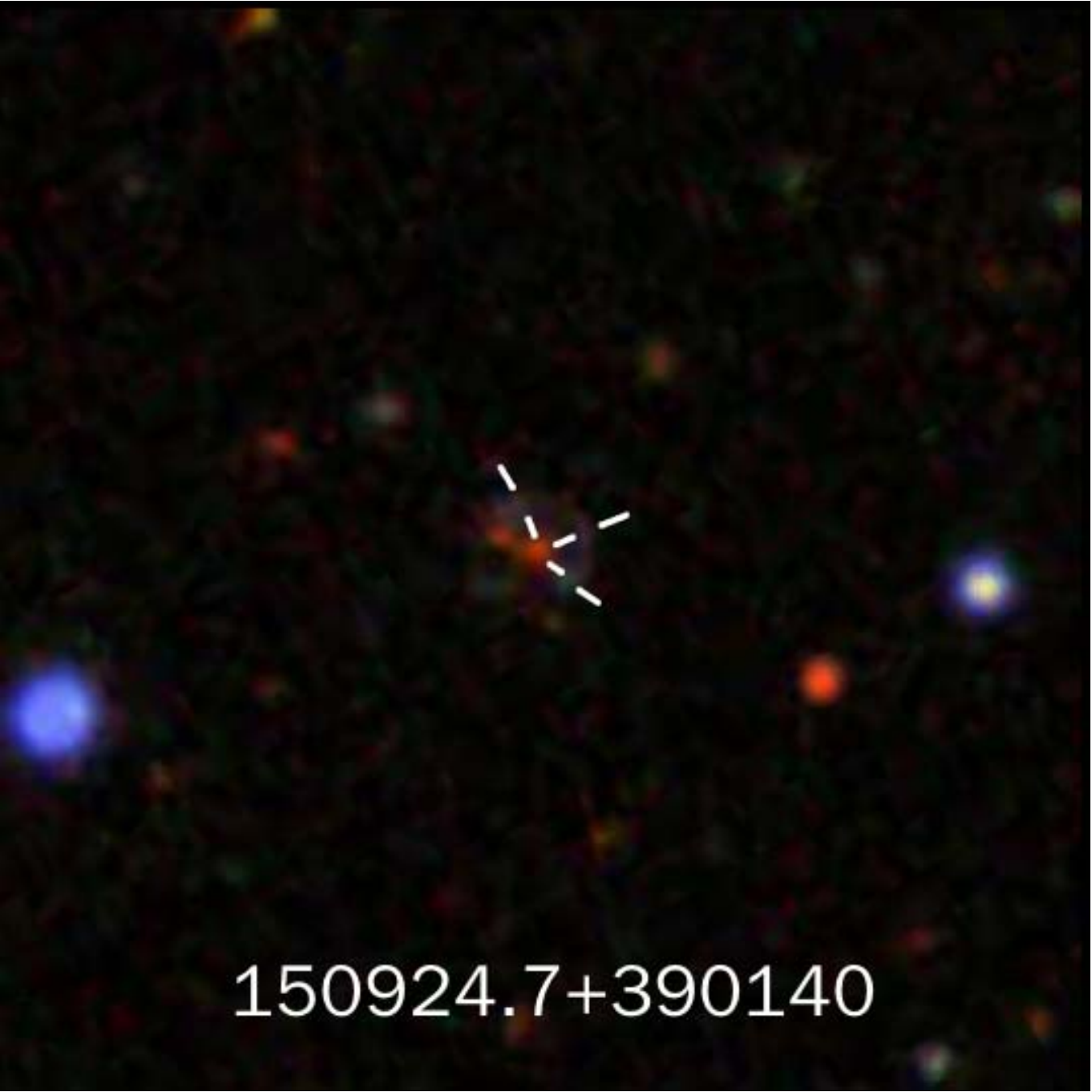}\\
\includegraphics[width = 2.1in]{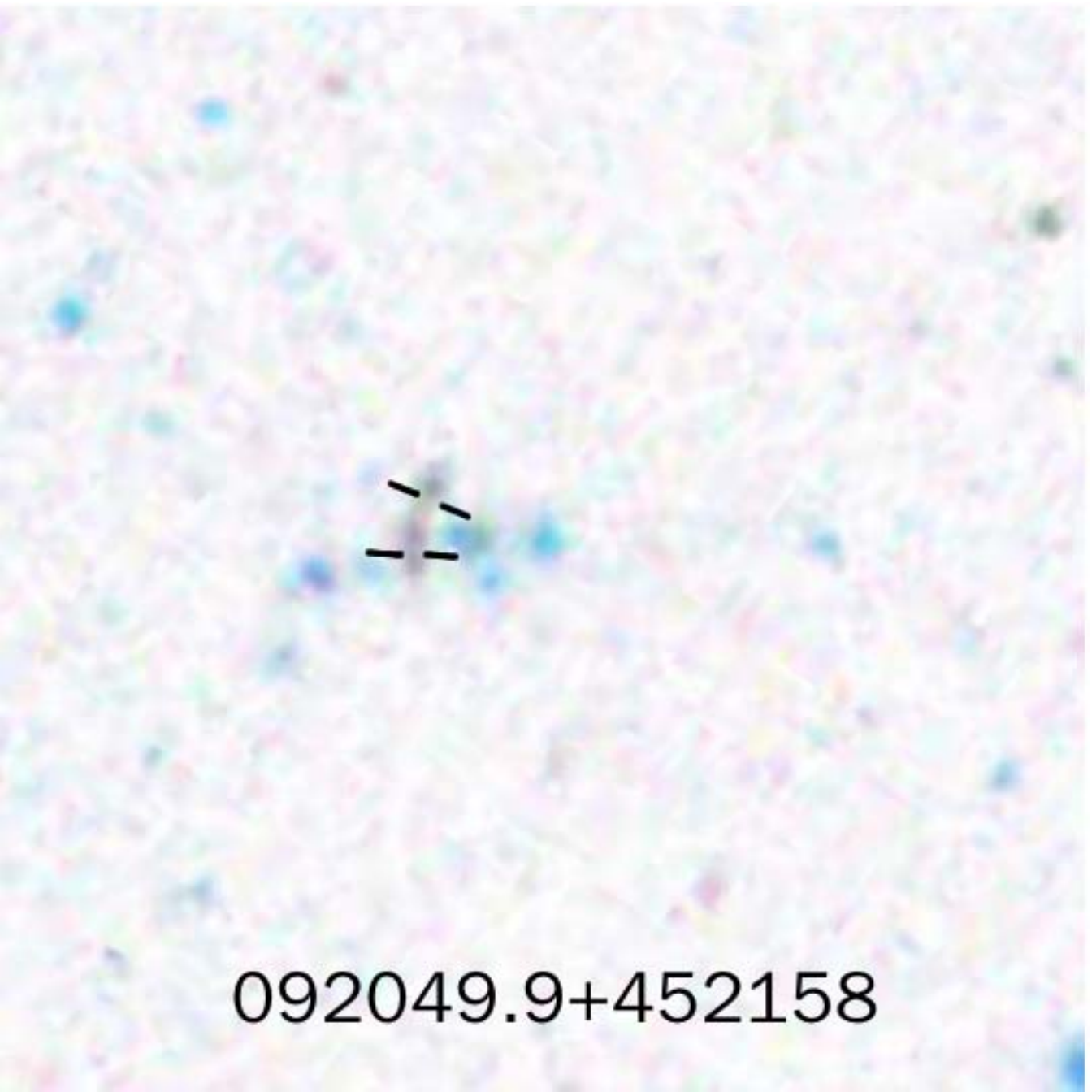}&
\includegraphics[width = 2.1in]{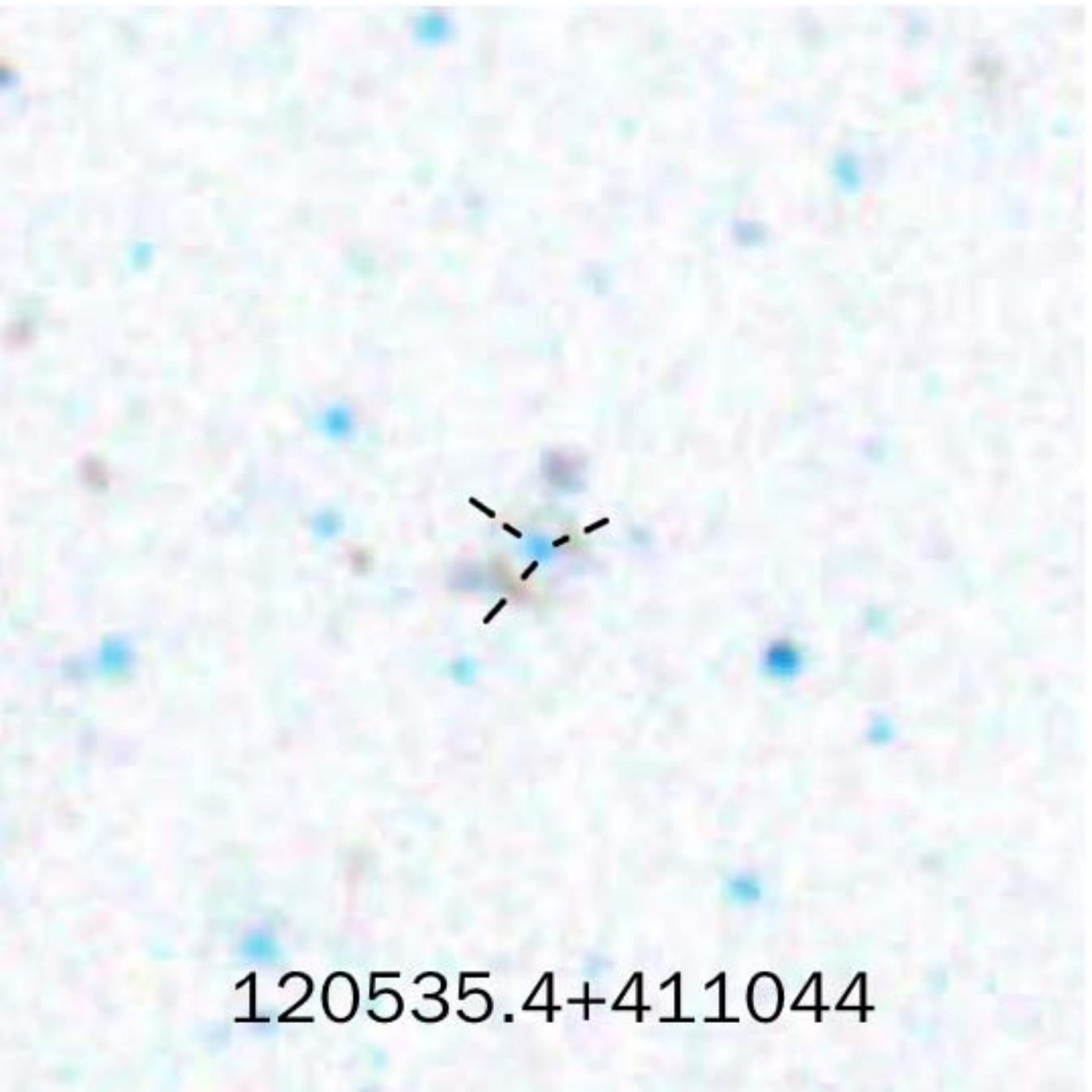}&
\includegraphics[width = 2.1in]{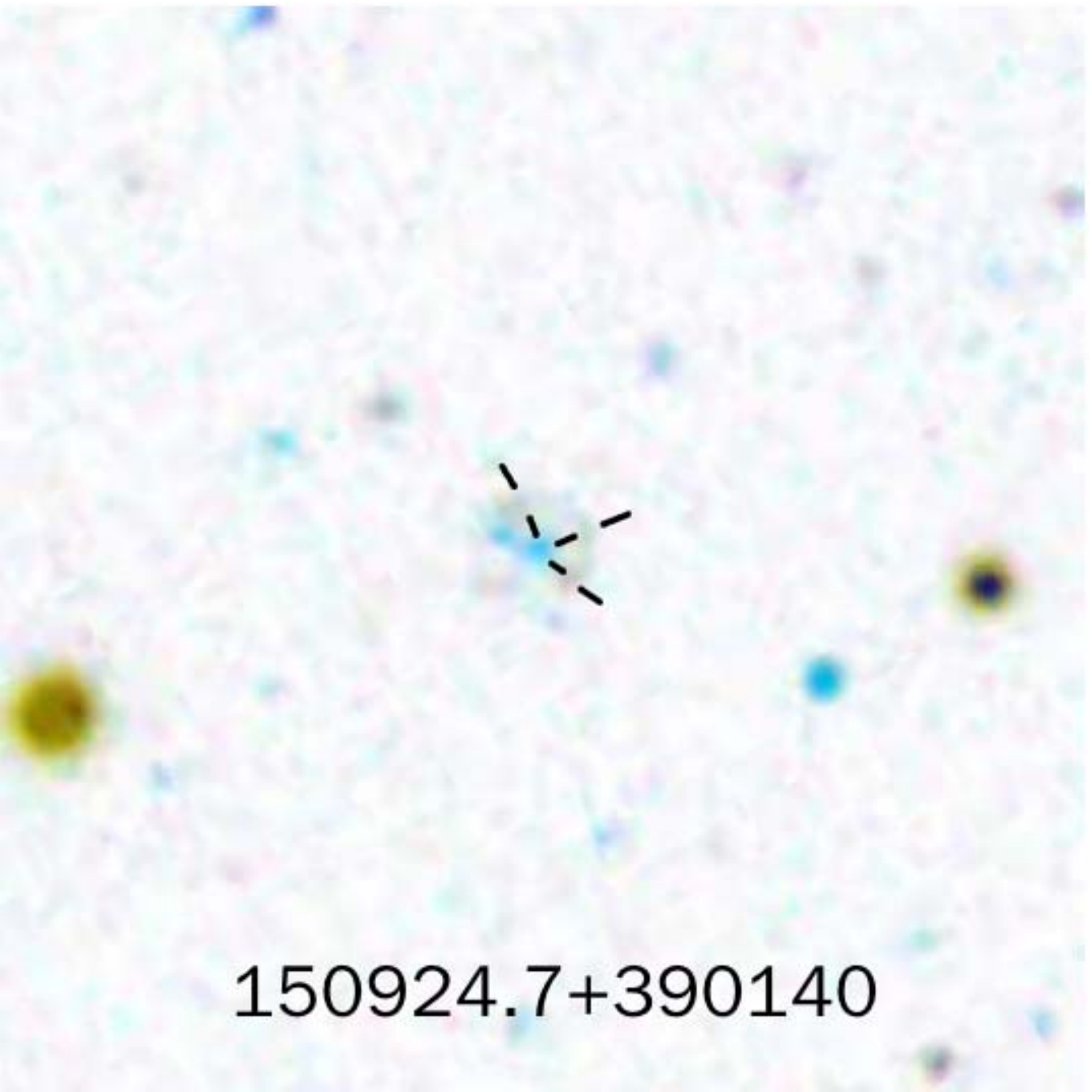}
\end{tabular}
\caption{SDSS images of 8 clusters with a field of view of
  1.2$'\times$1.2$'$.  They are almost certain to be gravitational lensing
  systems. The negative images are also shown in the second rows to
  see the lensing features more clearly.}
\label{almost}
\end{figure*}

\begin{figure*}
  \begin{tabular}{ccc}
\includegraphics[width = 2.1in]{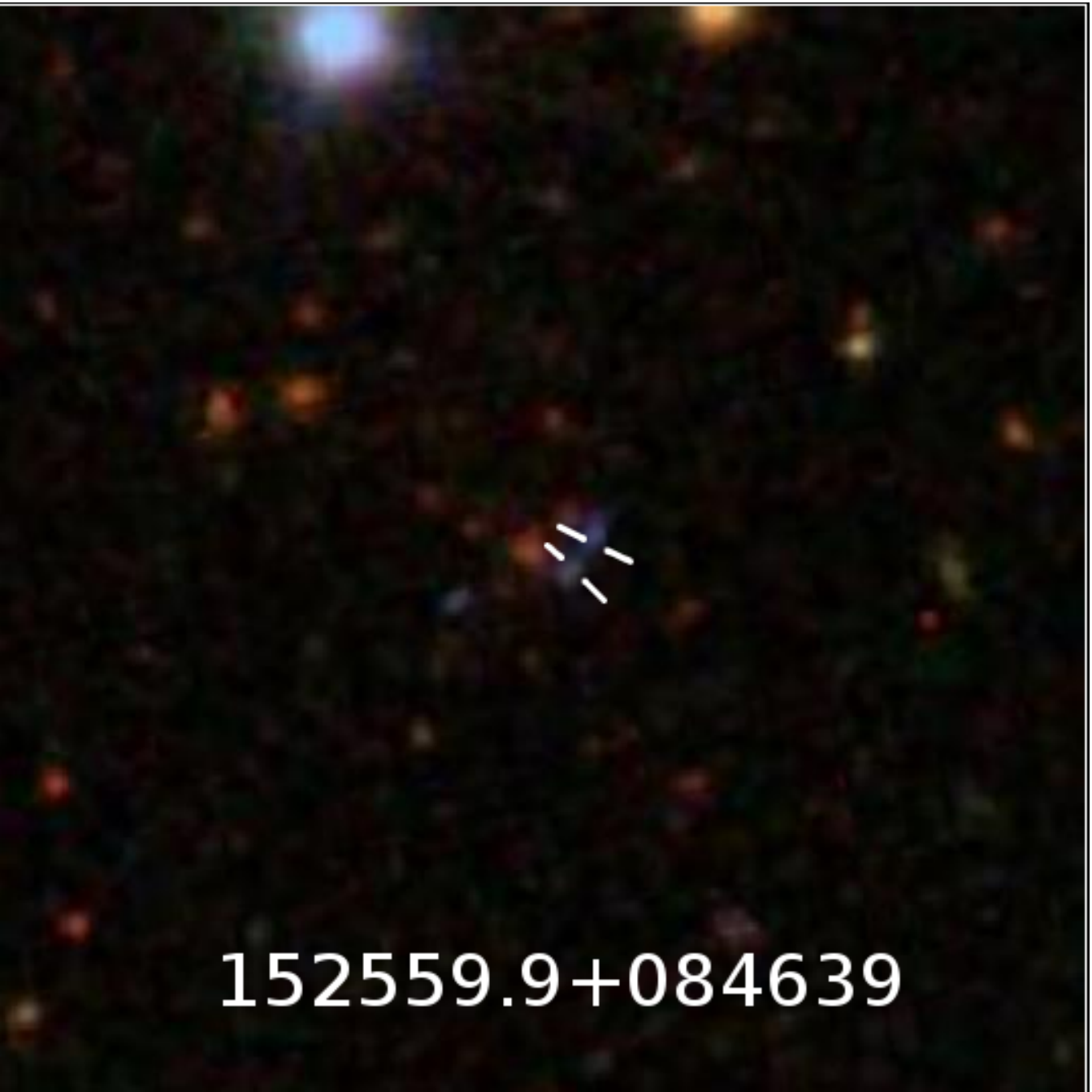}&
\includegraphics[width = 2.1in]{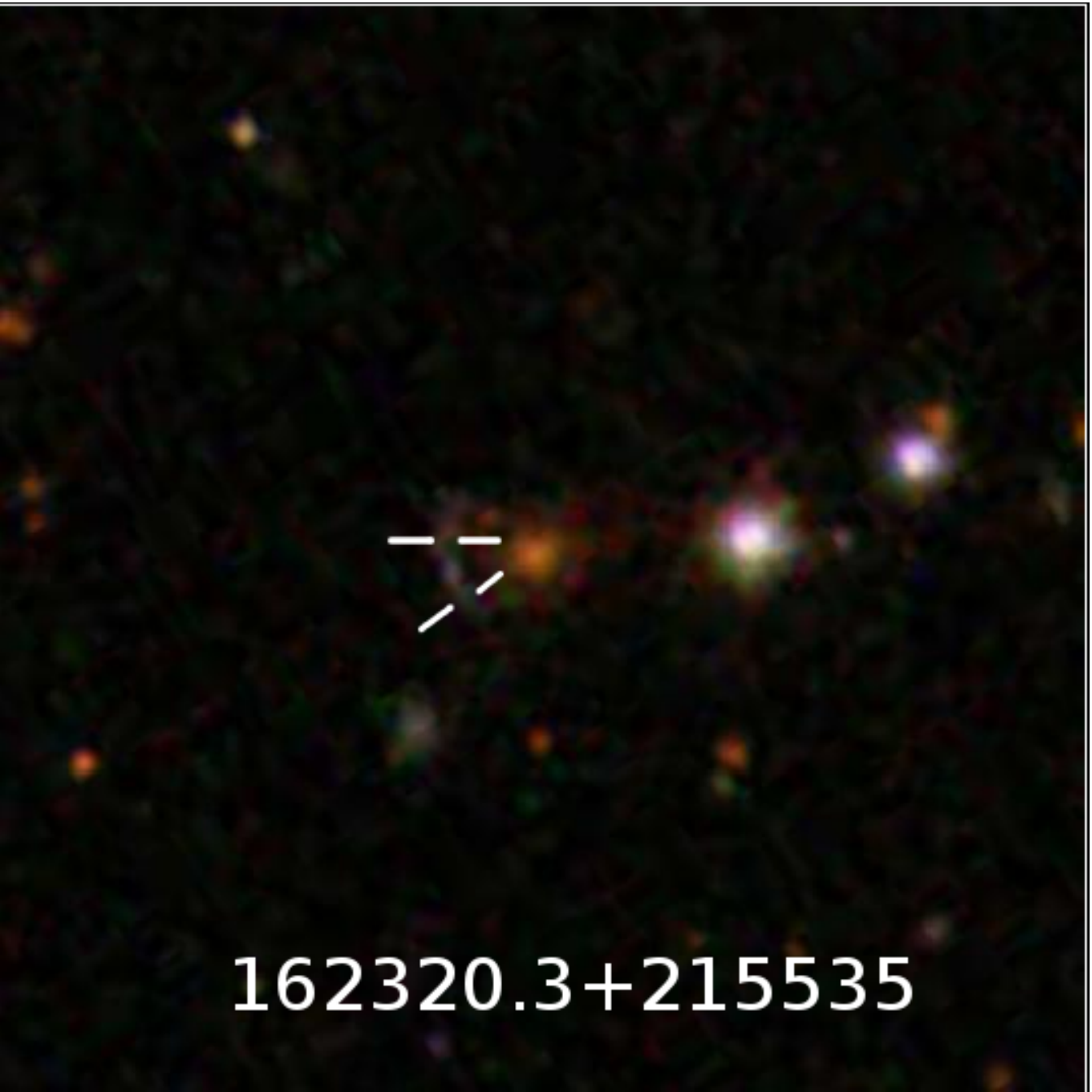}\\
\includegraphics[width = 2.1in]{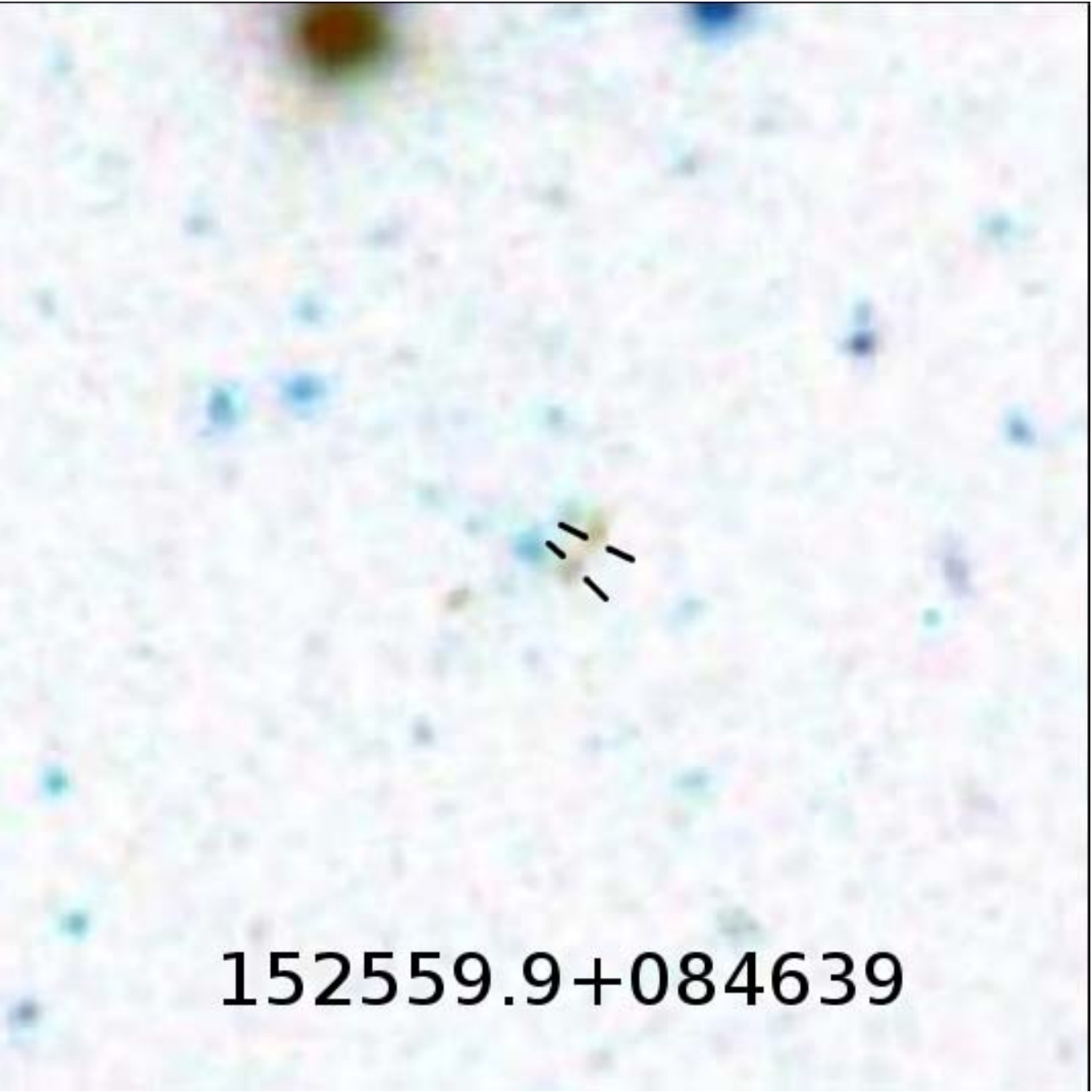}&
\includegraphics[width = 2.1in]{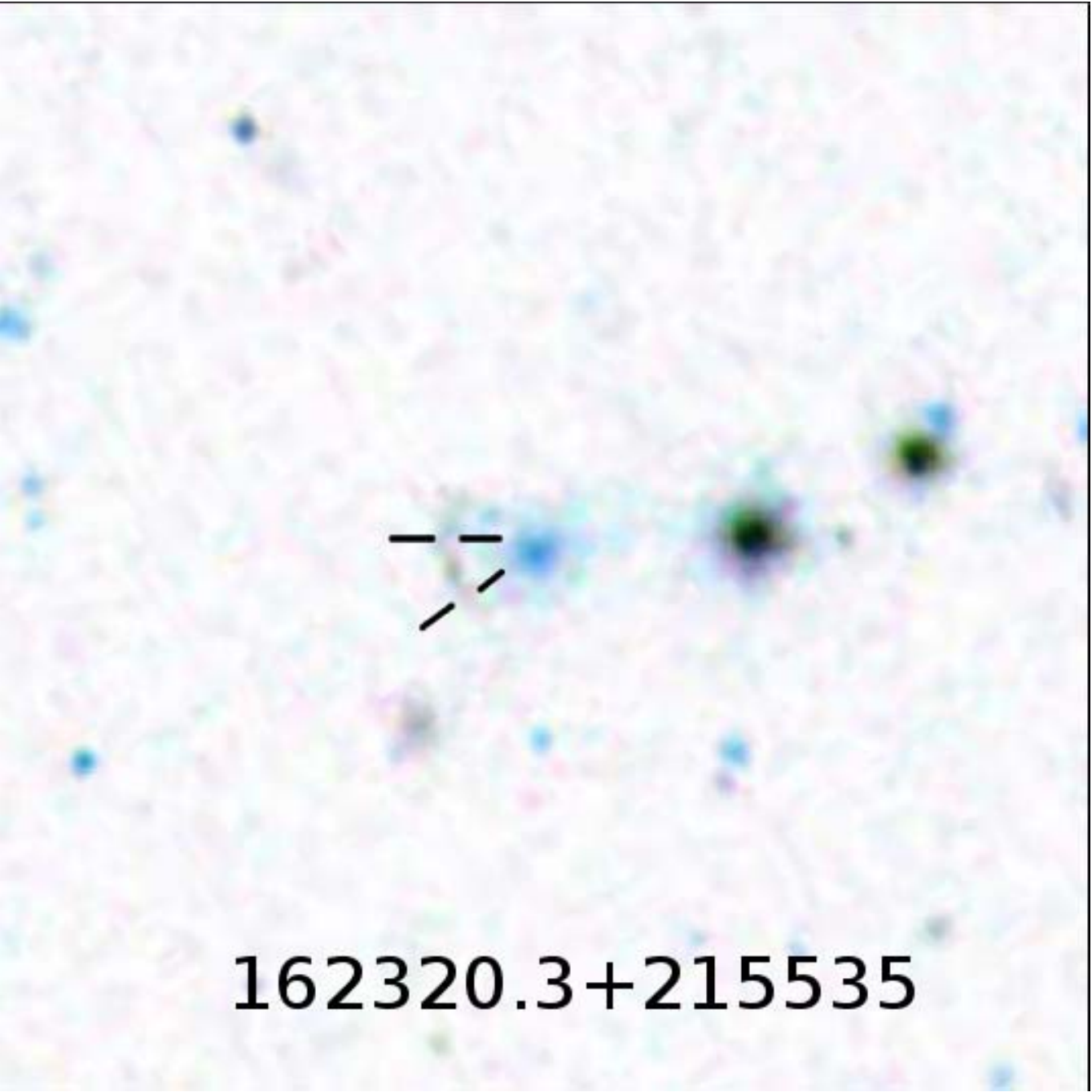}
\end{tabular}
\setcounter{figure}{0}
\caption{{\it continued}}
\end{figure*}

\begin{figure*}
\begin{tabular}{cccc}
\includegraphics[width = 2.1in]{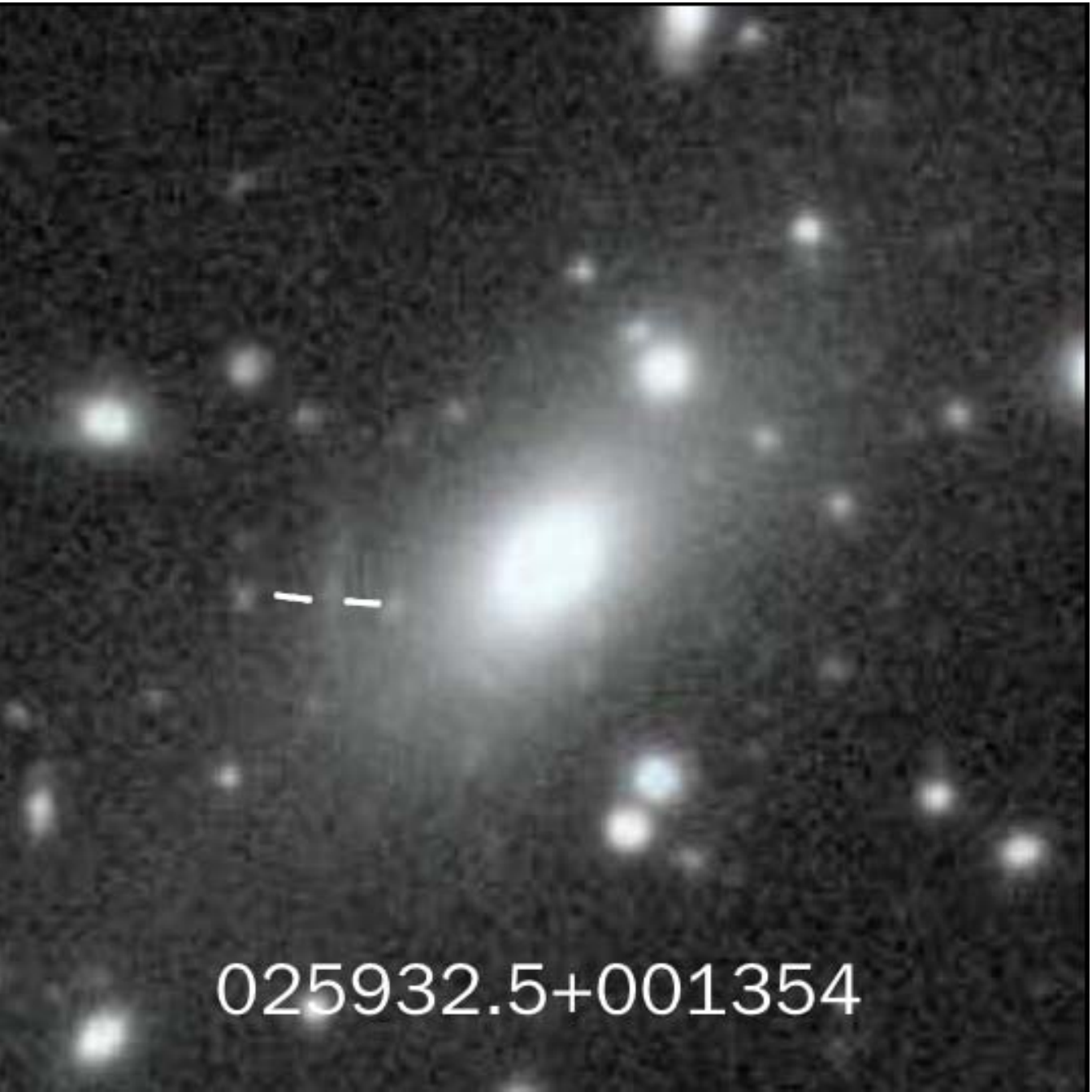}&
\includegraphics[width = 2.1in]{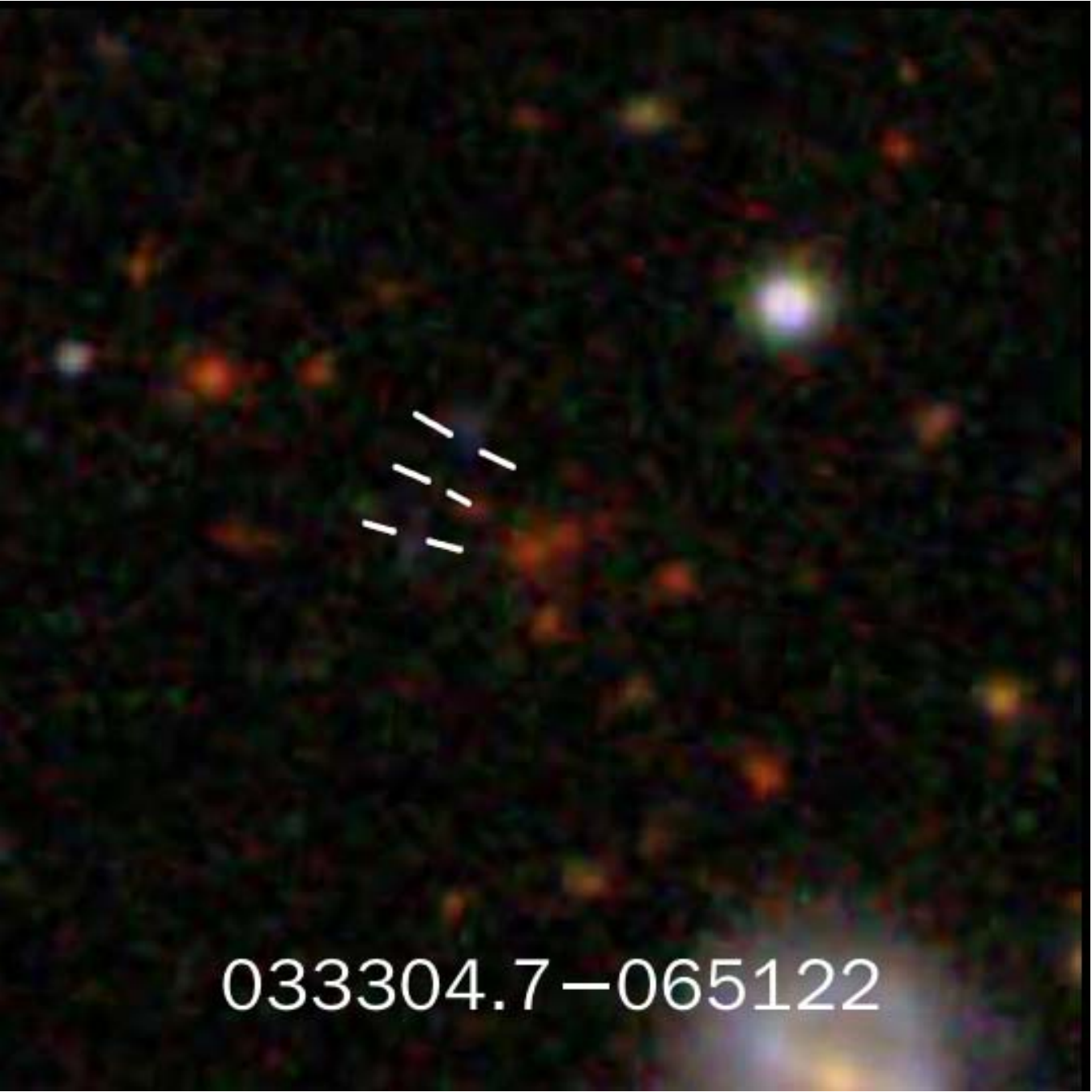}&
\includegraphics[width = 2.1in]{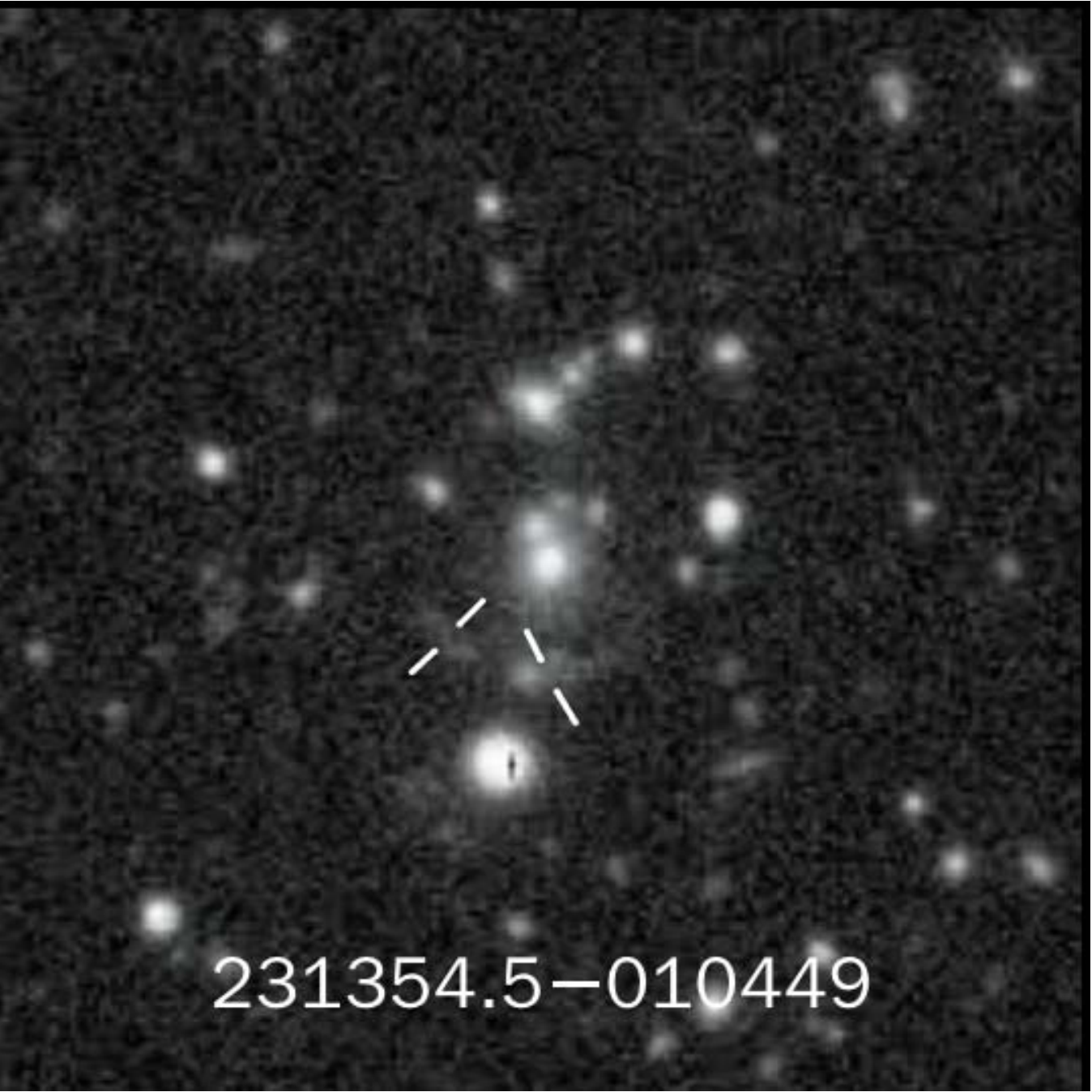}\\
\includegraphics[width = 2.1in]{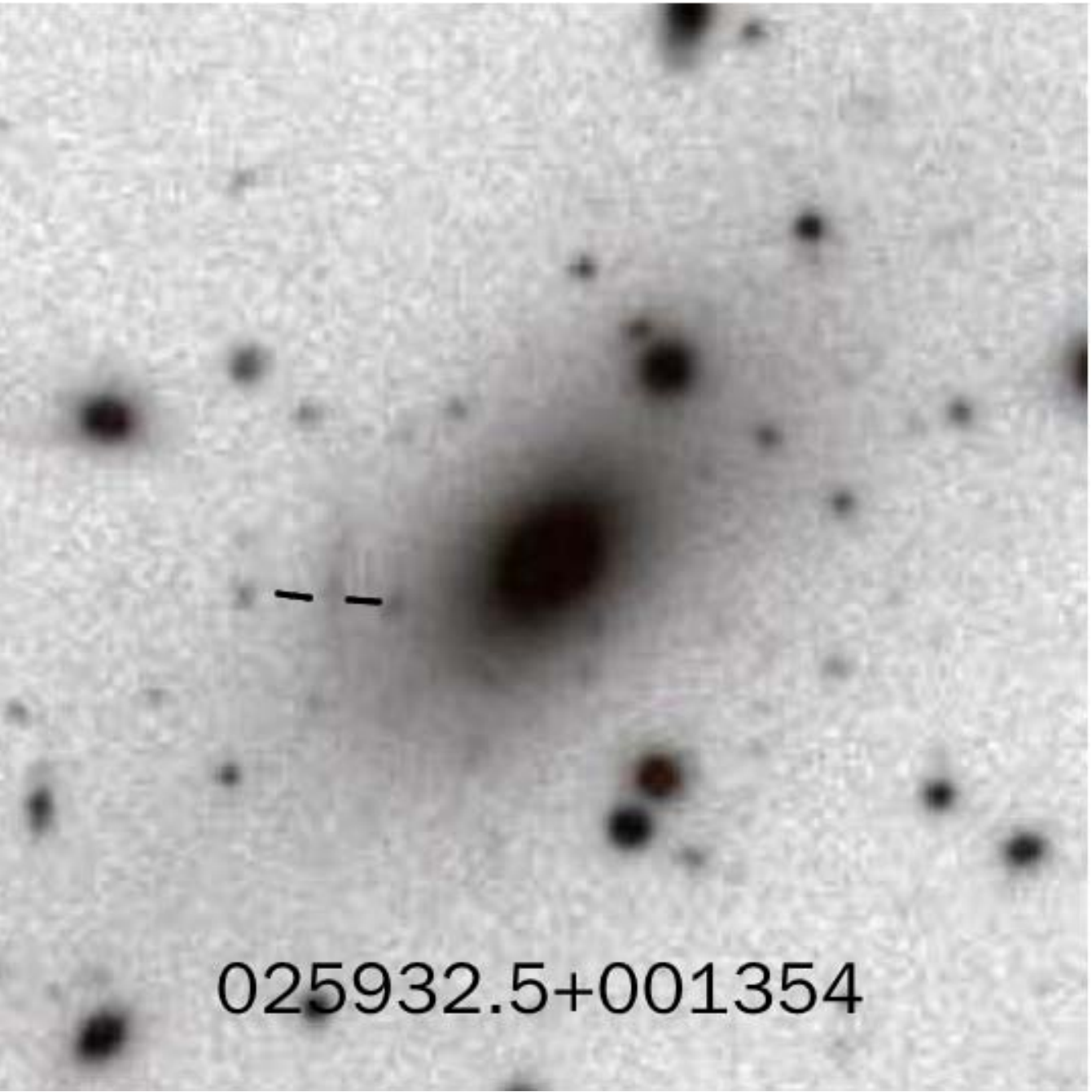}&
\includegraphics[width = 2.1in]{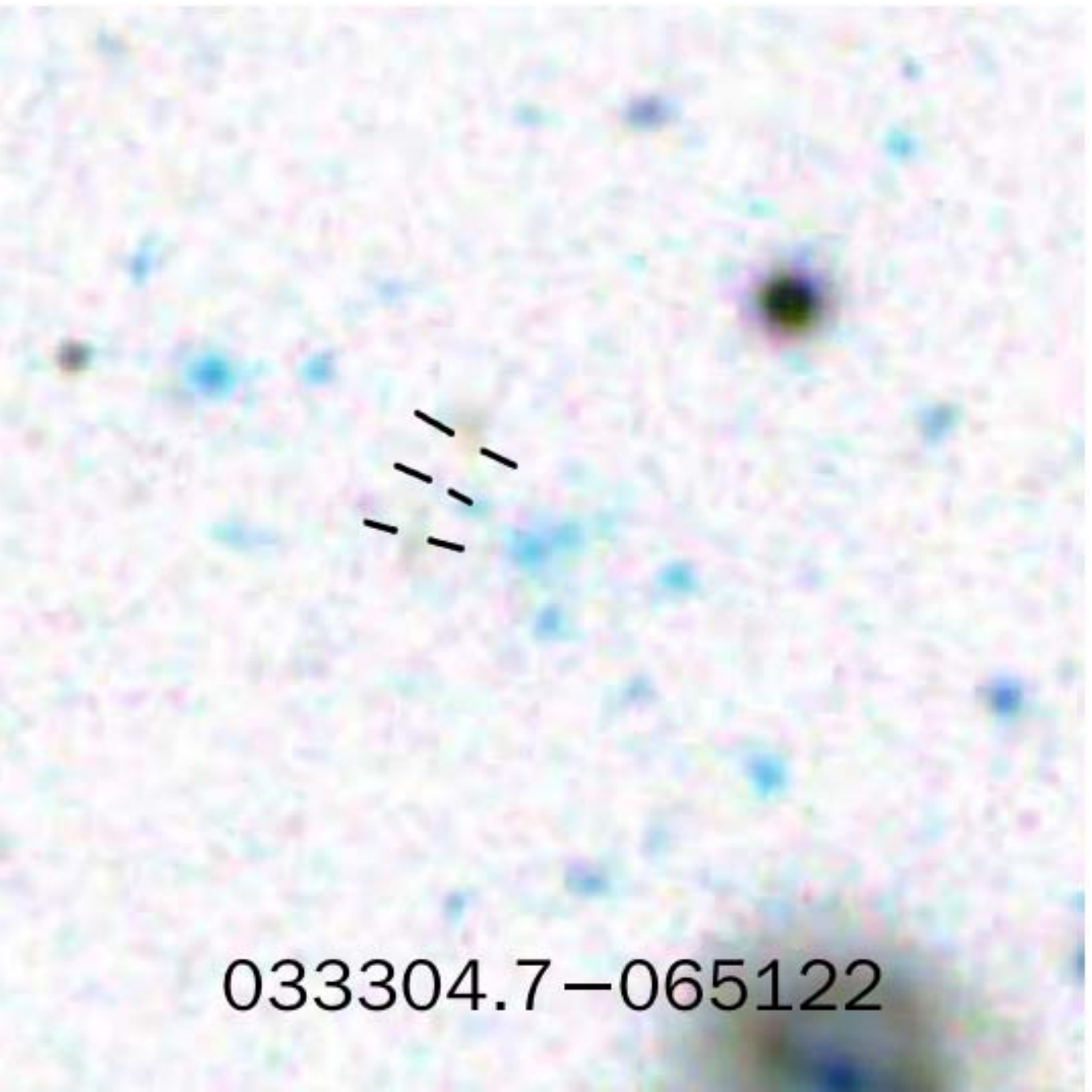}&
\includegraphics[width = 2.1in]{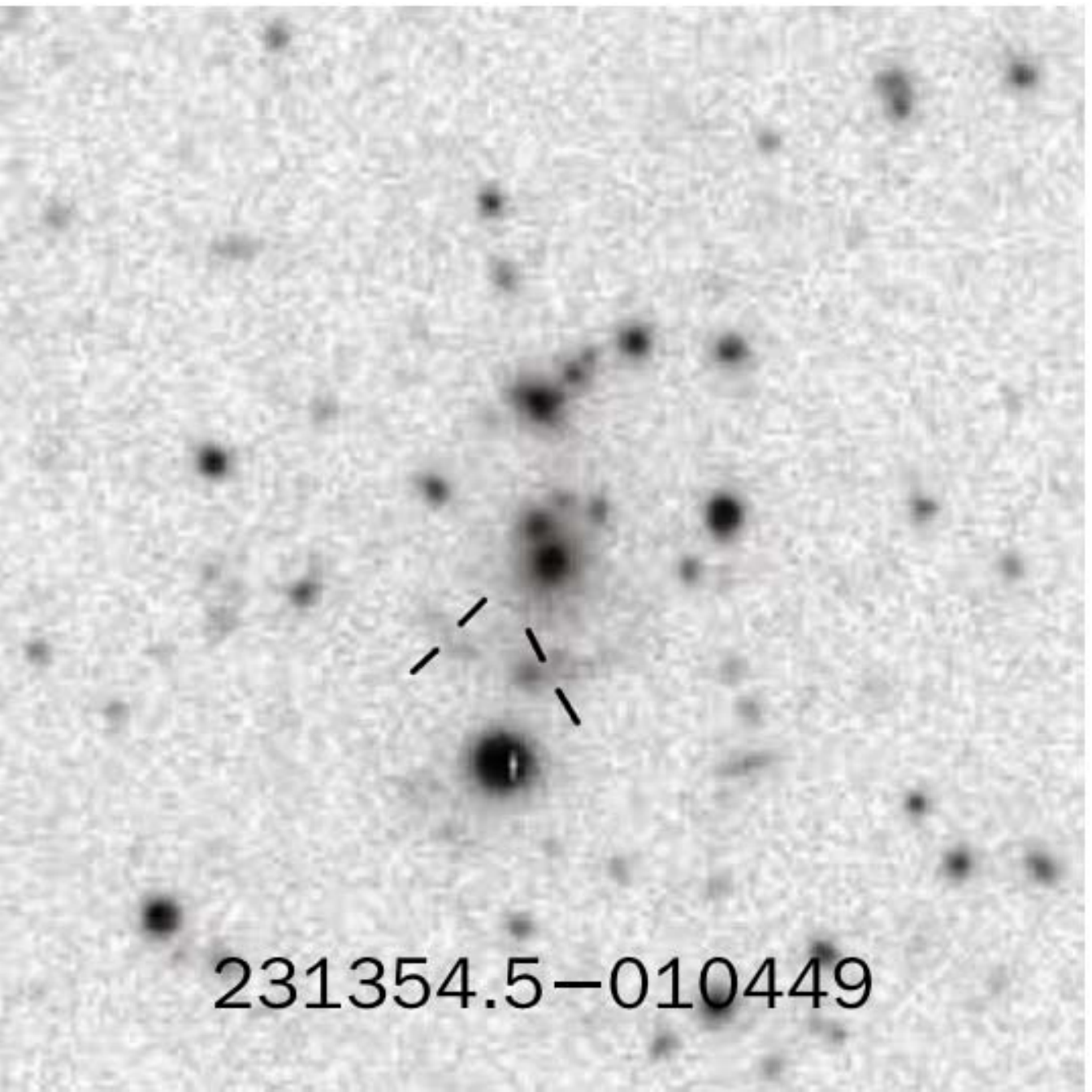}
\end{tabular}
\caption{SDSS images of 3 {\it probable} lensing clusters 
with a field of view of 1.2$'\times$1.2$'$. The negative images 
are also shown in the second rows.}
\label{proba}
\end{figure*}

\begin{figure*}
\begin{tabular}{lll}
\includegraphics[width = 2.1in]{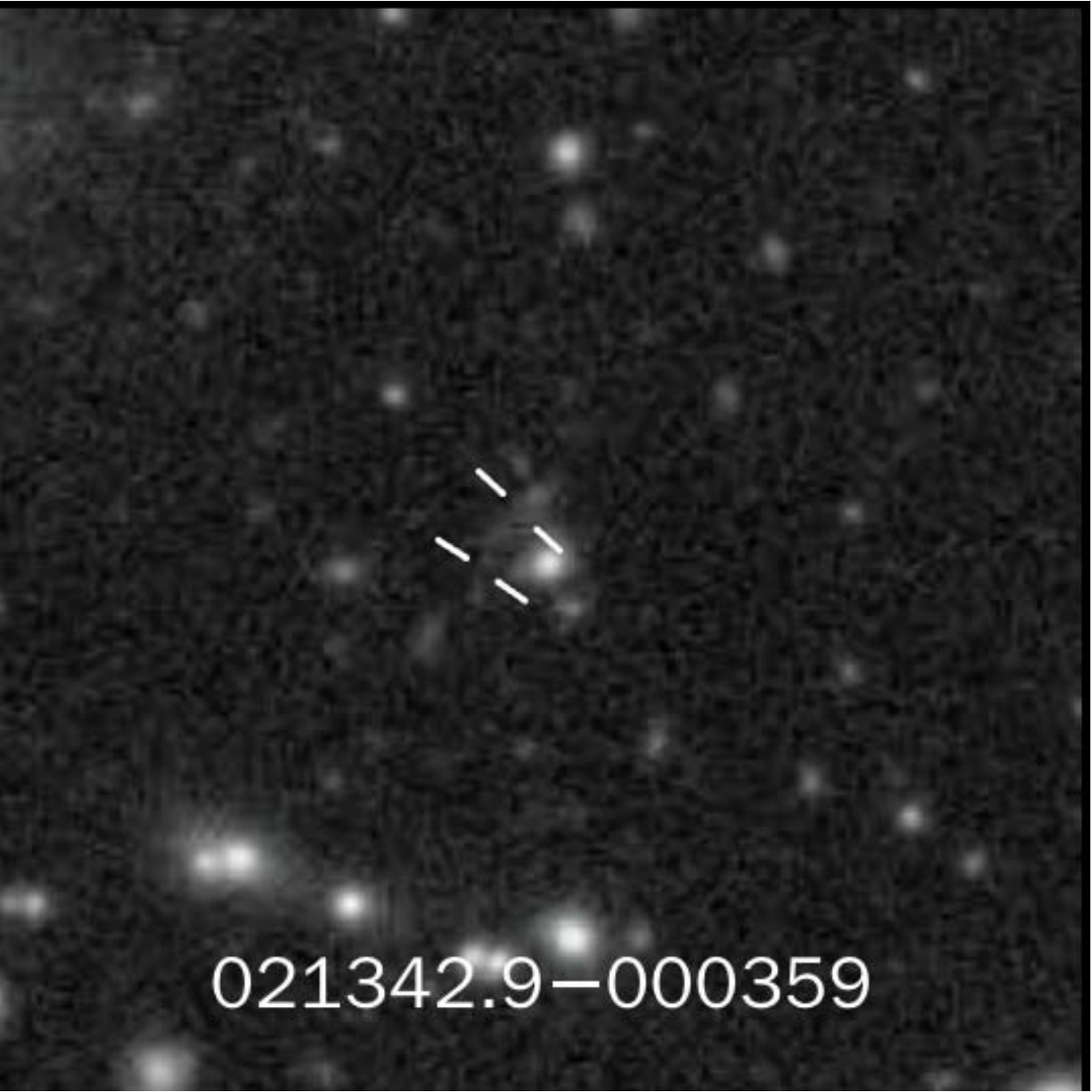}&
\includegraphics[width = 2.1in]{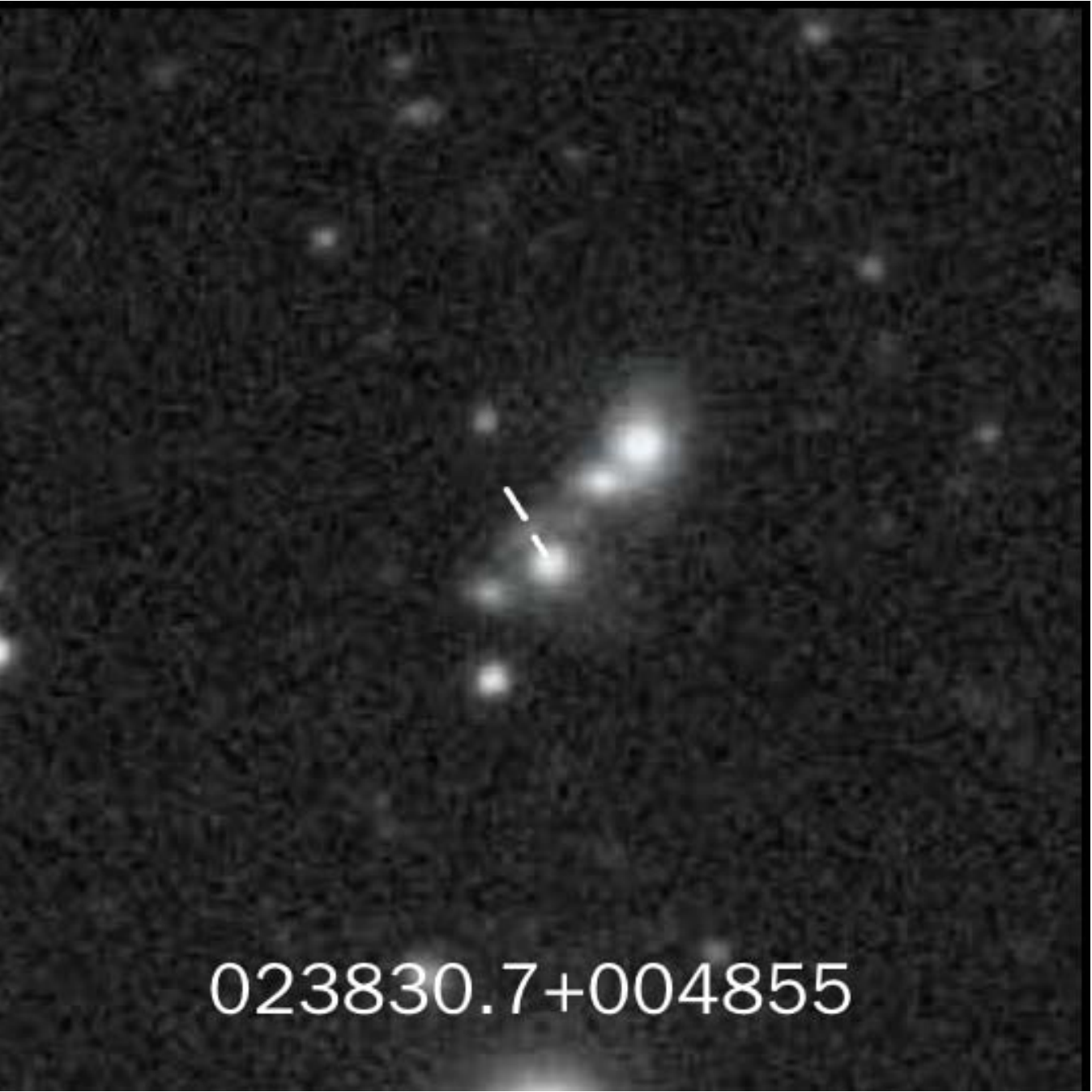}&
\includegraphics[width = 2.1in]{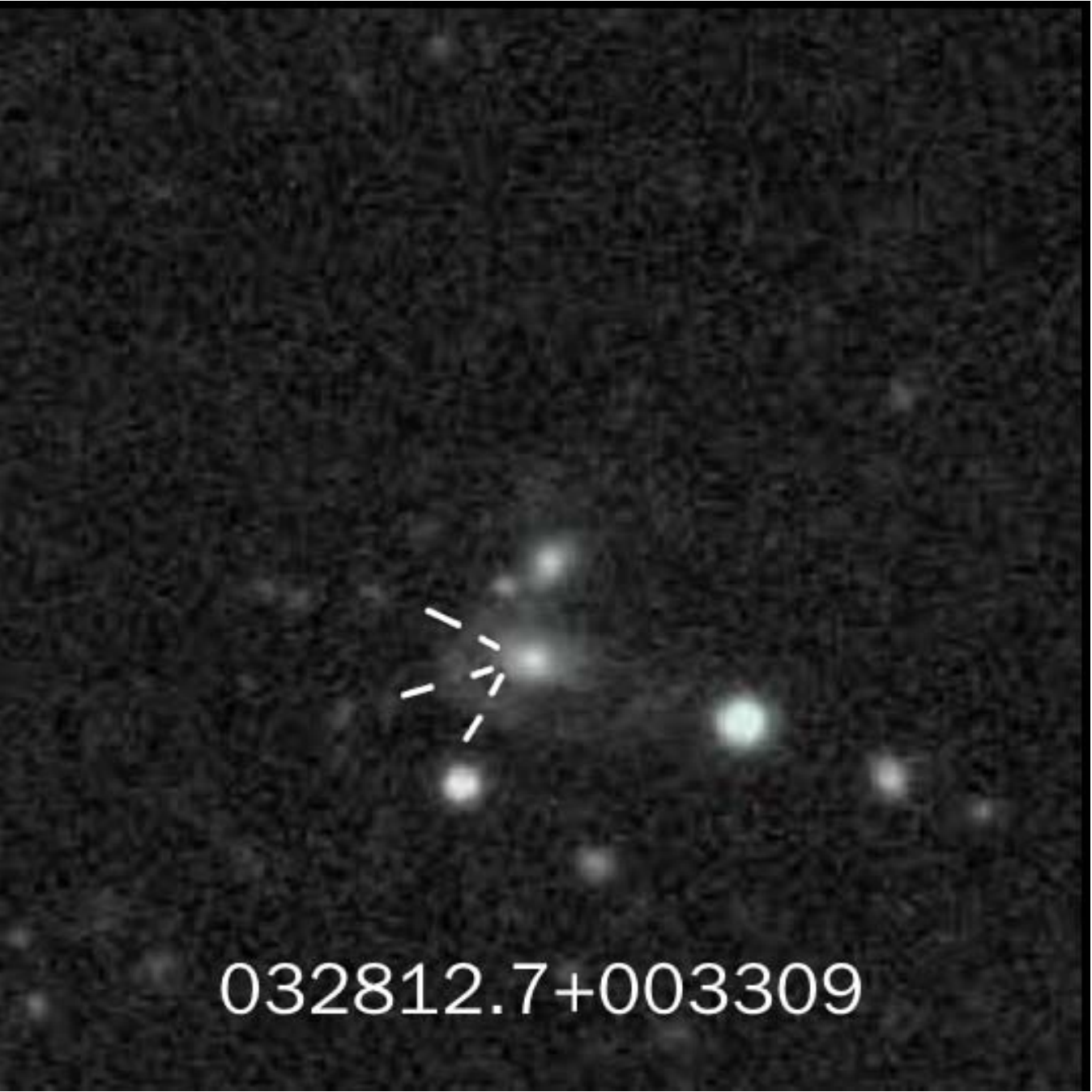}\\
\includegraphics[width = 2.1in]{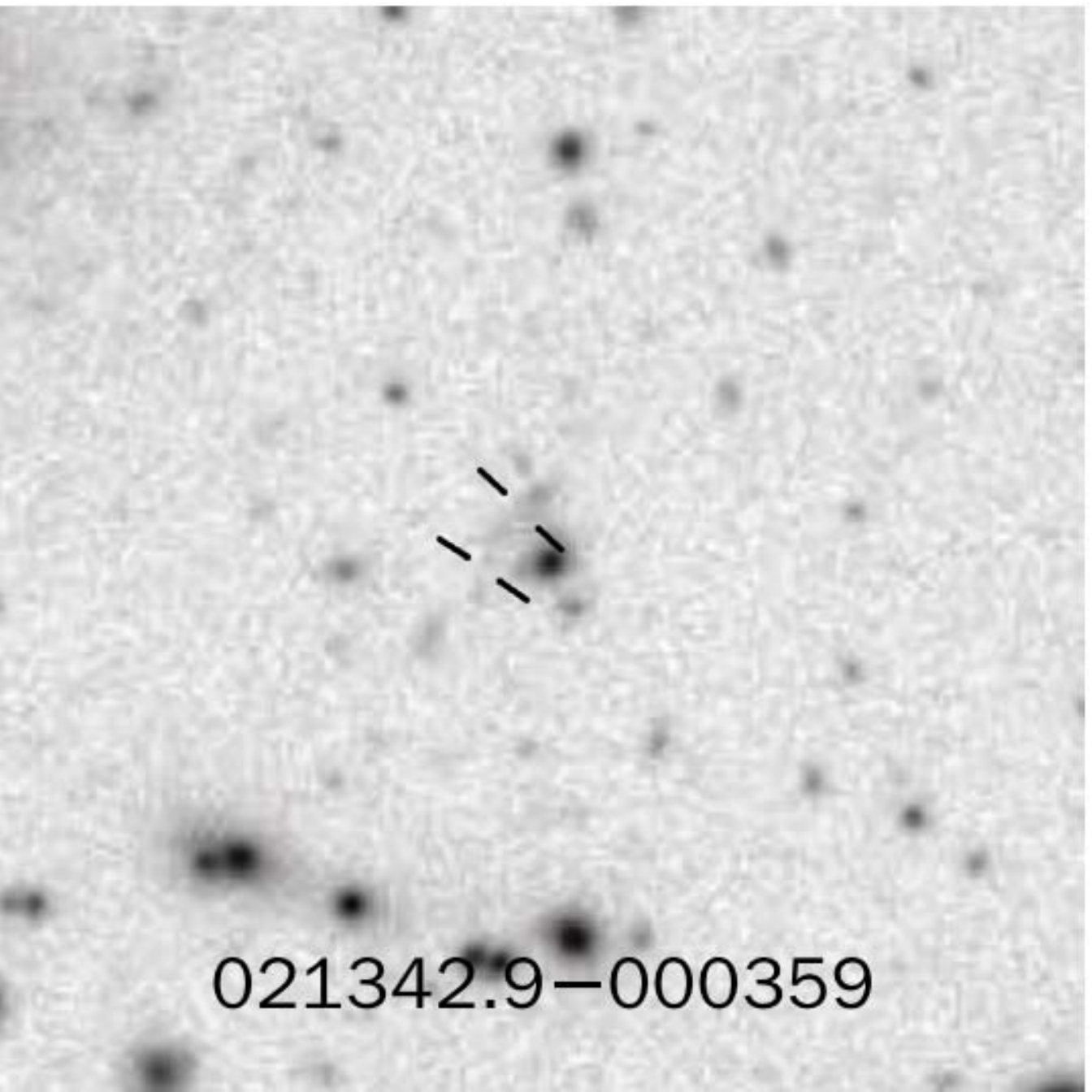}&
\includegraphics[width = 2.1in]{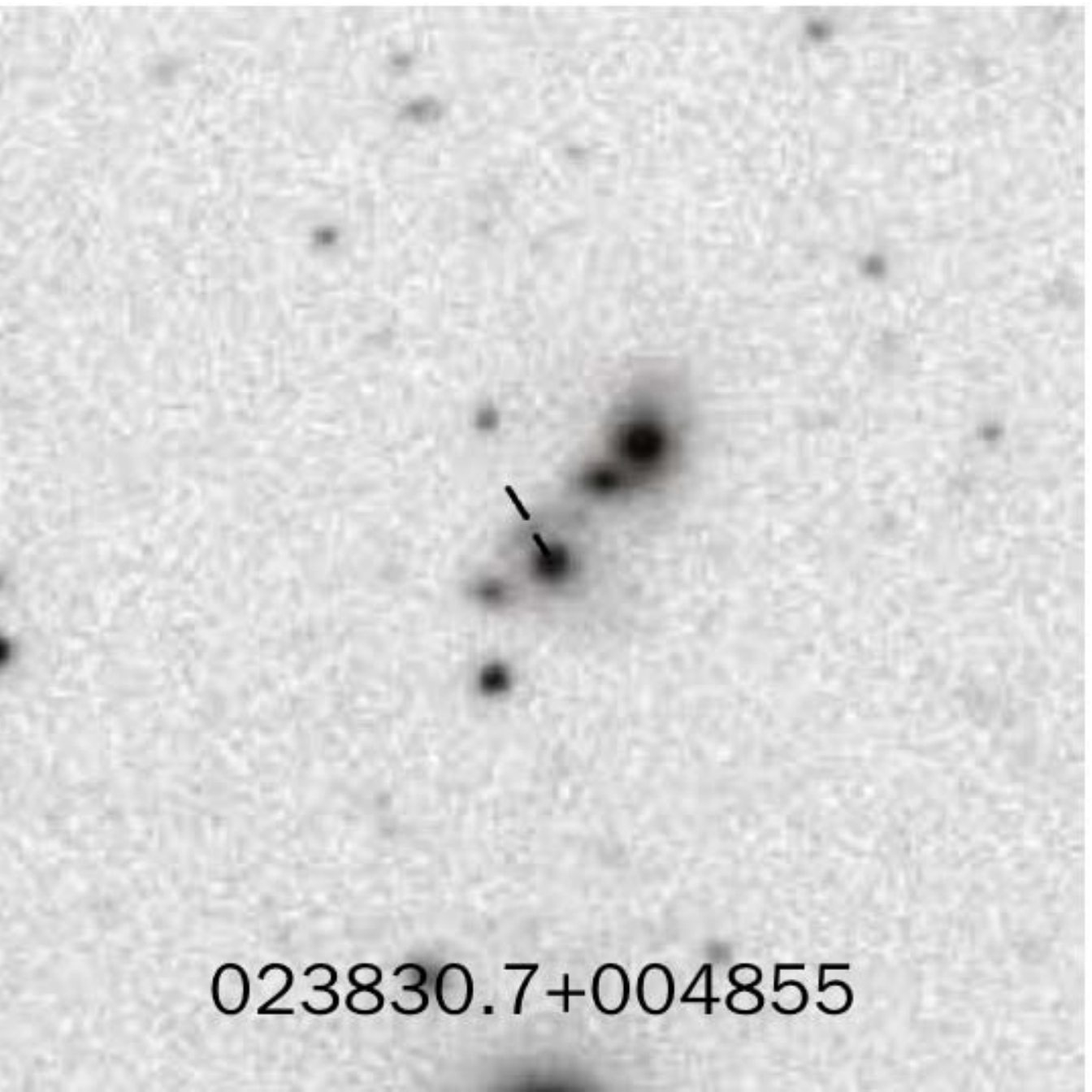}&
\includegraphics[width = 2.1in]{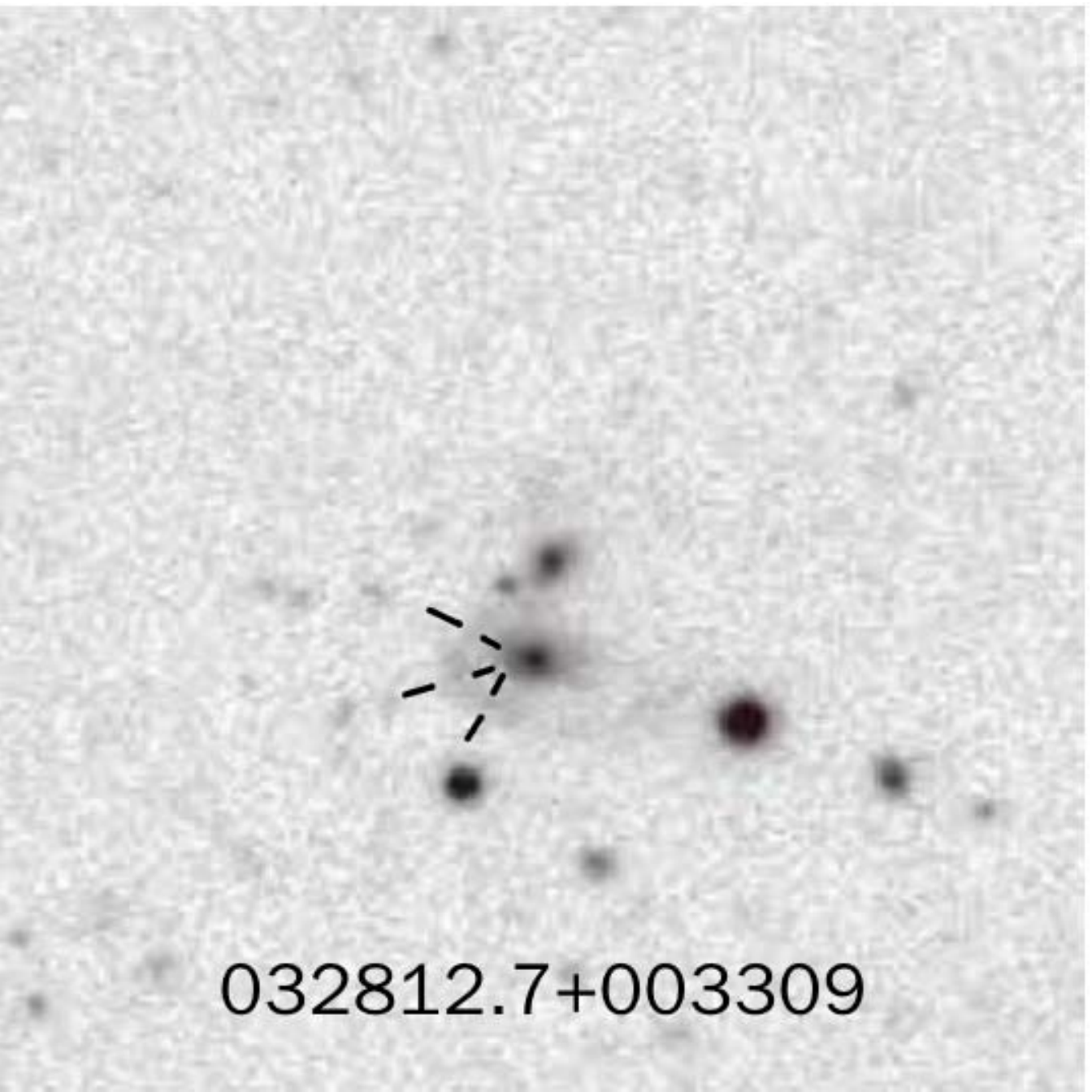}\\
\includegraphics[width = 2.1in]{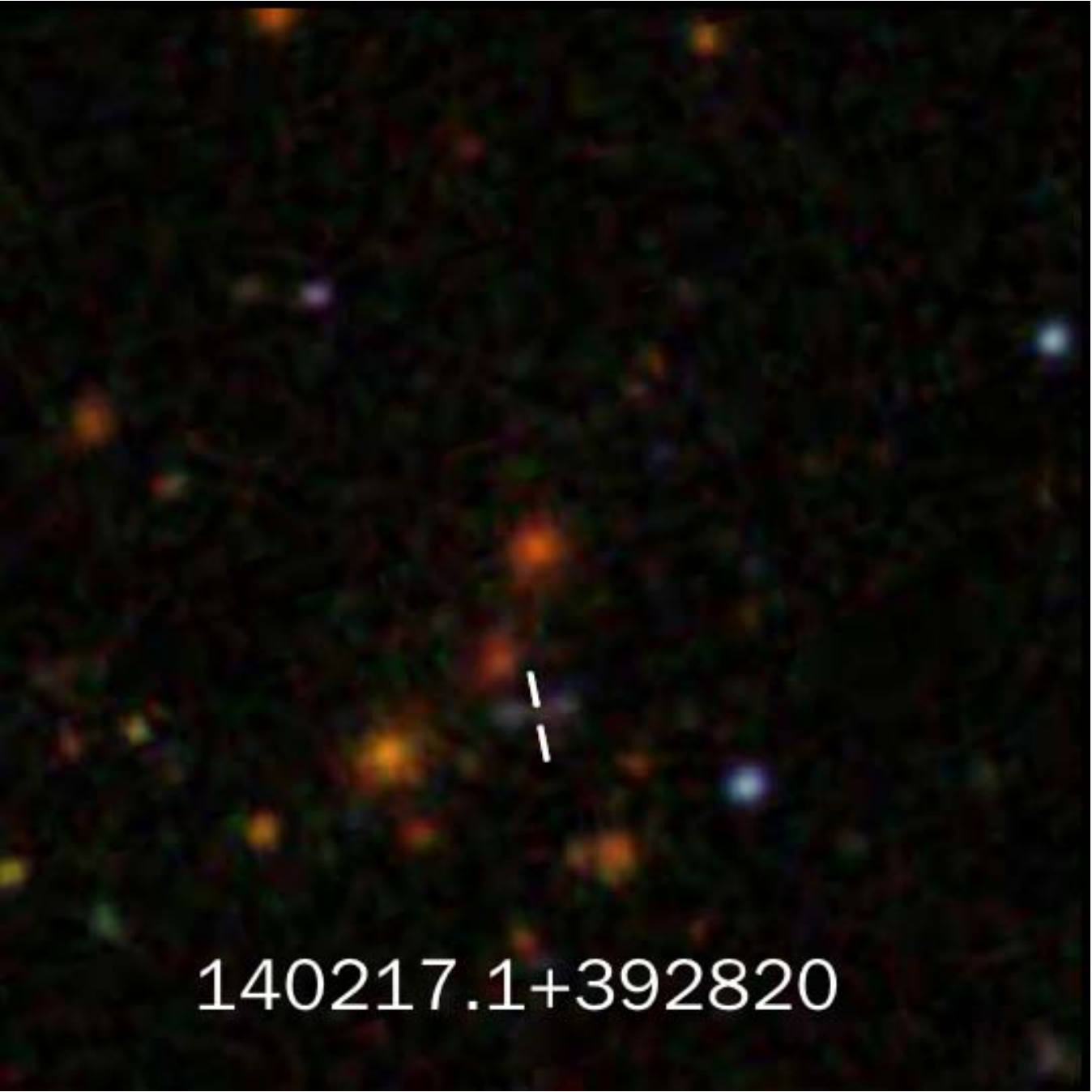}&
\includegraphics[width = 2.1in]{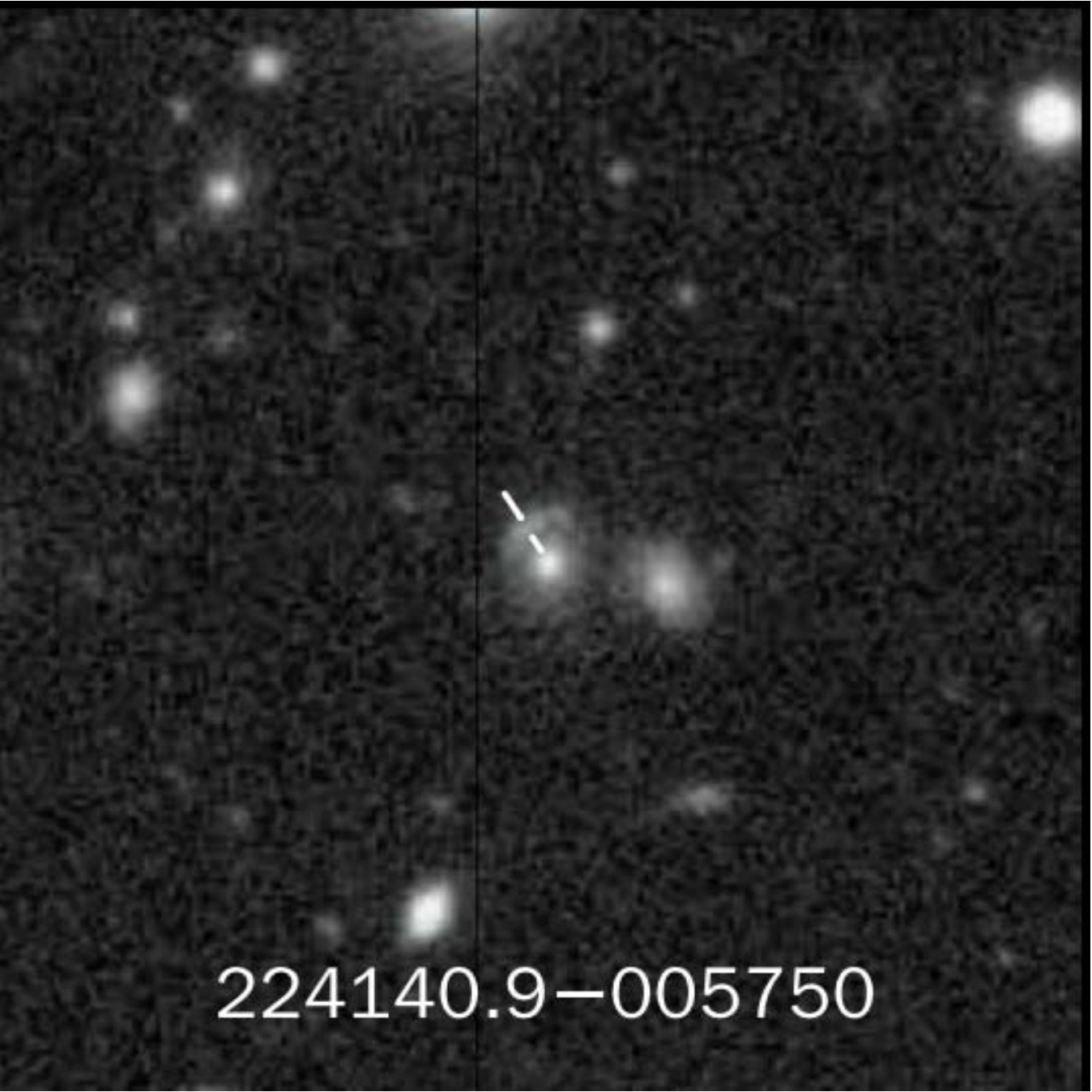}&
\includegraphics[width = 2.1in]{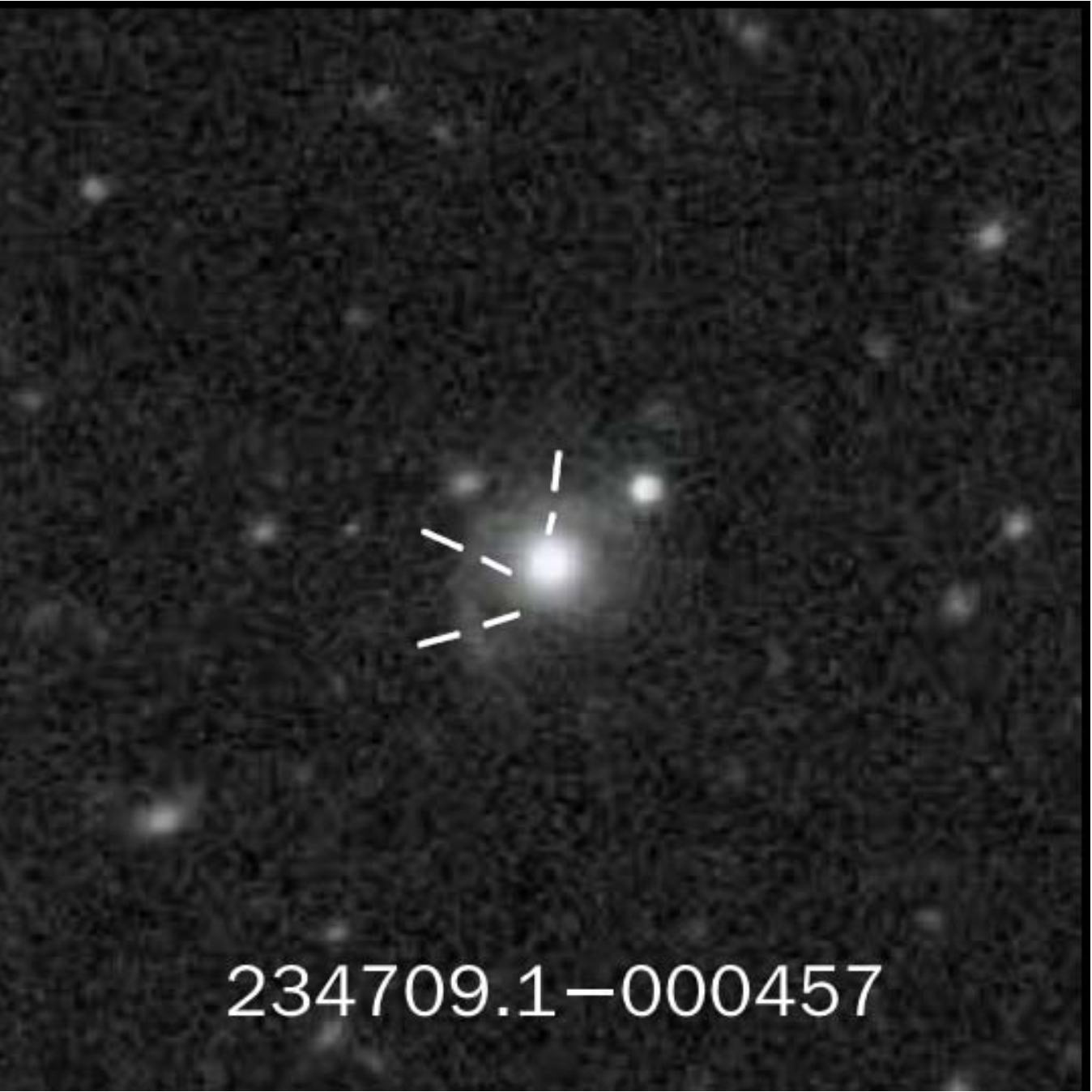}\\
\includegraphics[width = 2.1in]{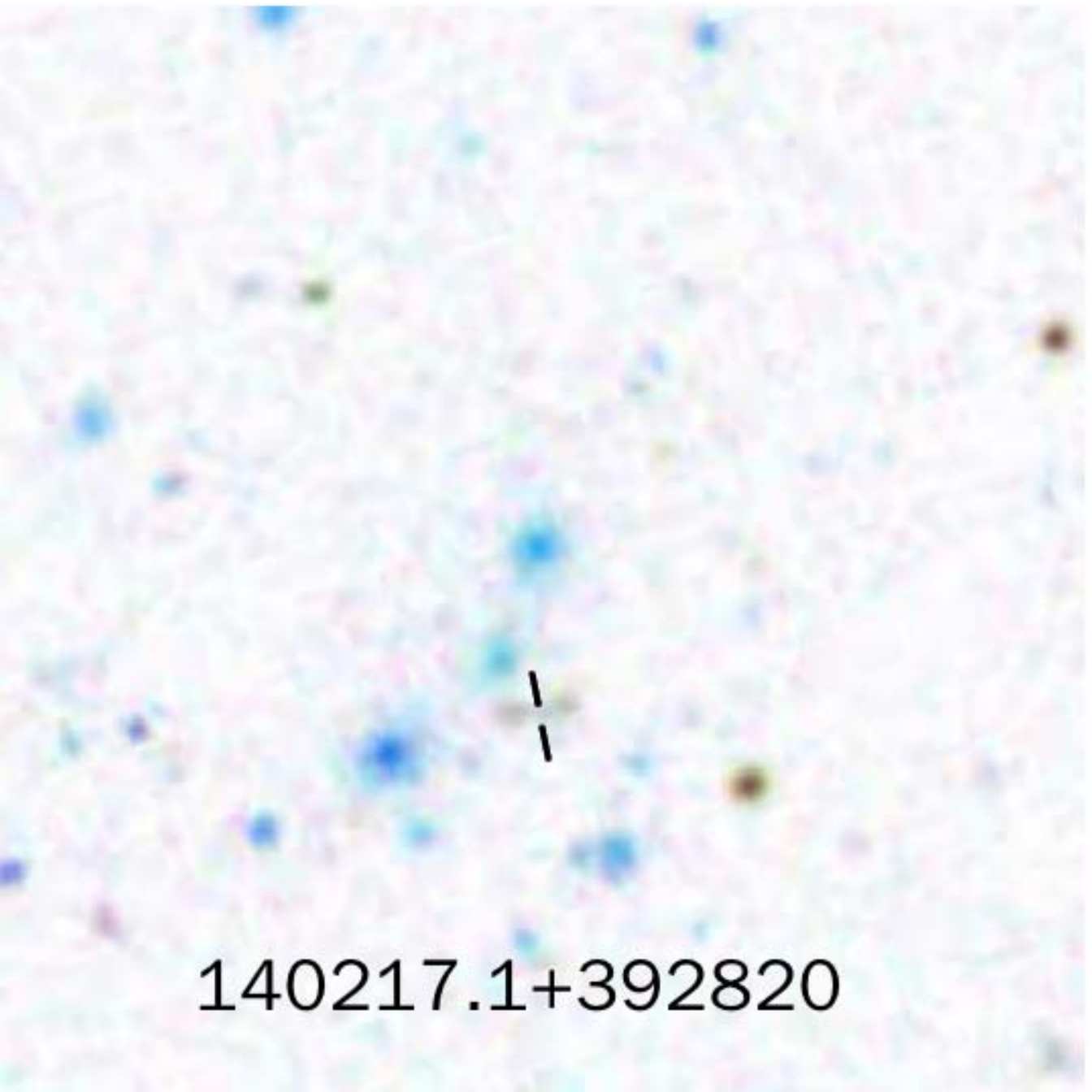}&
\includegraphics[width = 2.1in]{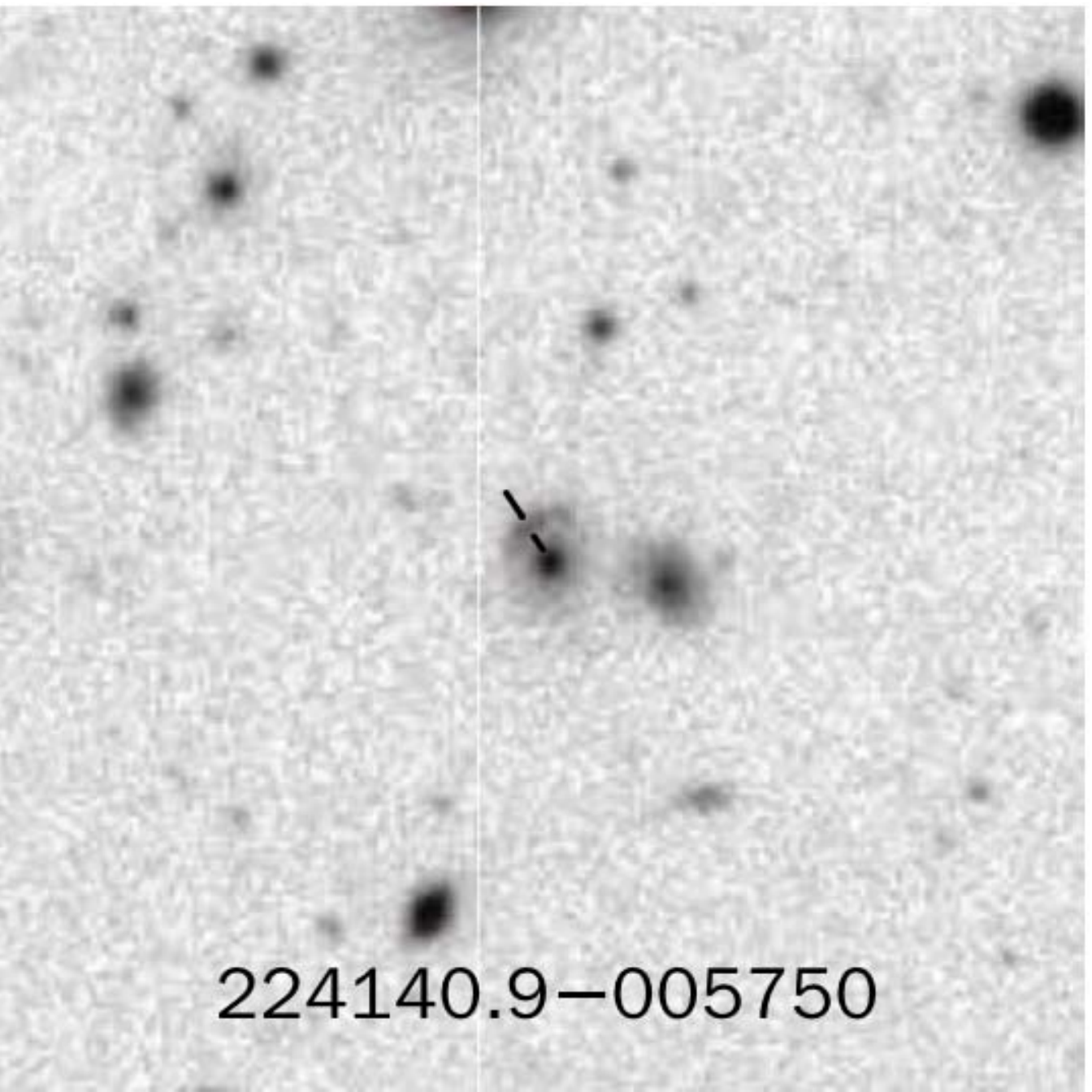}&
\includegraphics[width = 2.1in]{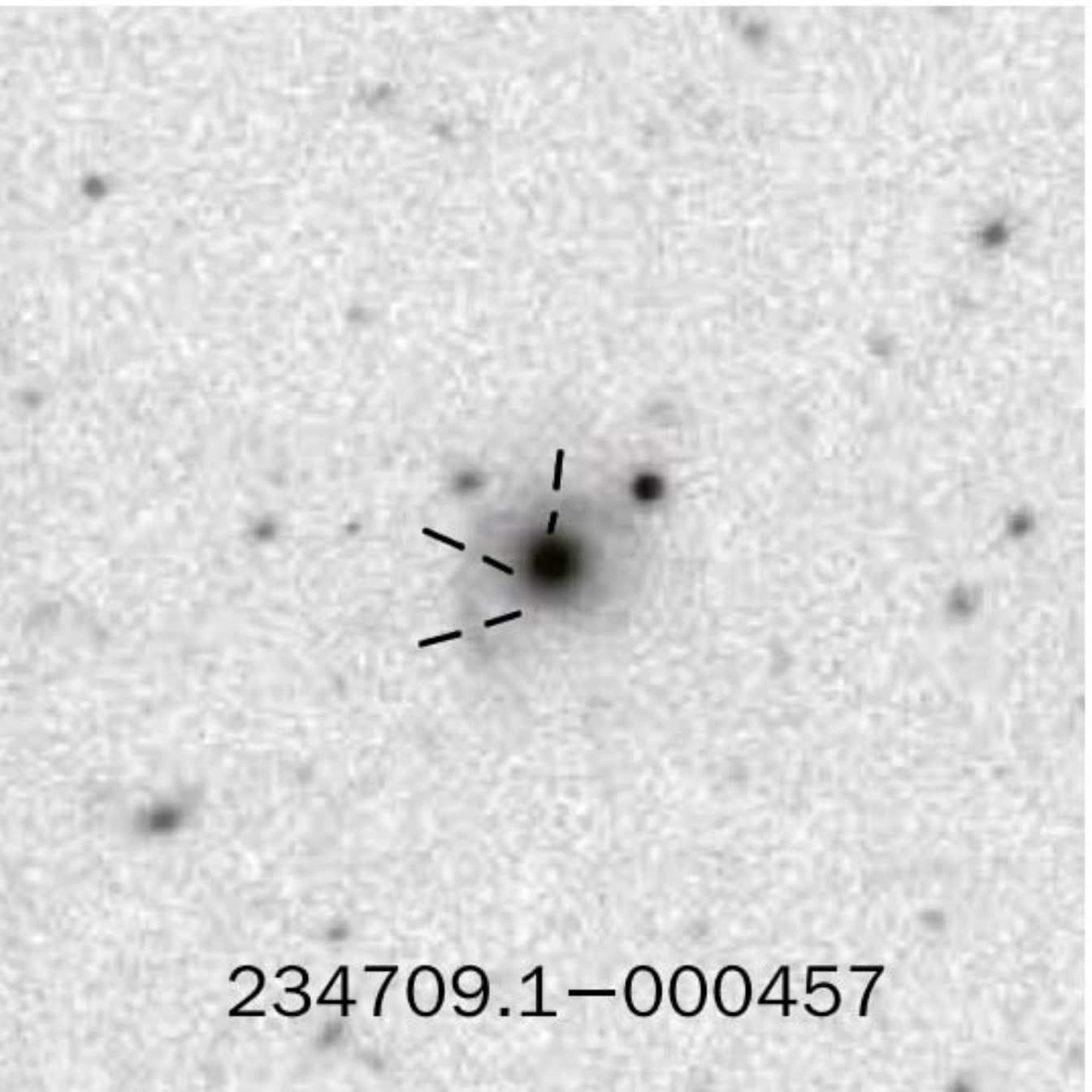}
\end{tabular}
\caption{SDSS images of 6 {\it possible} lensing clusters 
with a field of view of 1.2$'\times$1.2$'$. The negative 
images are also shown in the second rows.}
\label{possi}
\end{figure*}

Using the SDSS data, we identified two large samples of galaxy
clusters, and then visually inspected the color images of the clusters
to find new lensing systems. Our first effort was made by inspecting
the color images of 39,668 clusters, from which we found 13 new
lensing systems candidates [10]. They were classified as 4 almost
certain lensing systems, 5 probably and 4 possible cases. Currently,
12 of these 13 systems have been confirmed as true lensing systems by
spectroscopic observations [13--15]. The second effort was to view
color images of 132,684 clusters, from which 98 lensing candidates
were found [11] . They were classified as 13 almost certain lensing
systems, 22 probably and 31 possible cases, and 2 exotic
systems. Among them, 16 candidates have been confirmed [16]. We found
from this large sample that richer clusters have a higher probability
to be lensing systems.

This paper is the third effort on the direction. Following our
previous procedures [17--18], we first identify new high redshift
clusters from the SDSS DR8 data and the SDSS Stripe 82 data, then
check images of 66,033 cluster candidates newly identified from the
SDSS DR8 and 46,663 candidates in the Stripe 82, and we find 17
candidates of strong lensing systems, among which 8 cases are almost
very certain, and other 3 are probably and 6 are possible cases. In
the following section, we introduce data of our galaxy cluster sample
and present our results of lensing searches. We then discuss the
redshift distribution, and estimated the masses of galaxy clusters
interior to the arcs, and then discuss the mass-to-light ratio for
galaxy clusters.

\sec{2\quad Data and Results}

Searches for new strong lensing systems are critical for both
individual and statistical studies of galaxy clusters. The probability
for strong lensing is very low. Currently, only a few hundreds among 
hundred thousands of galaxy clusters are shown as lensing systems.

We recently identify a new sample of galaxy clusters at high redshifts
from the SDSS DR8 and the Stripe 82. The SDSS survey data contain the
photometric redshifts of galaxies with a uncertainty of 0.025--0.030
in the redshift range of $z<0.45$ and with a larger at higher
redshift, by which large cluster samples were obtained [17--18]. A
cluster was recognized when the richness $R_{L*}=L_{\rm
  total}/L^{\ast}\ge12$ and the number of member galaxy candidates
within a photometric redshift slice of $z\pm0.04(1+z)$ and a radius of
$r_{200}$, $N_{200}\ge8$ [18]. Here, $r_{200}$ is the
radius within which the mean density of a cluster is 200 times of the
critical density of the universe, $L_{\rm total}$ is the total
luminosity of member galaxies, $L^{\ast}$ is the characteristic
luminosity. Most of the identified clusters have a redshift of $z<0.5$
because of the larger uncertainties of photometric redshift at higher
redshift. To find more high redshift clusters, we adjust the
photometric redshift slice to be $z\pm0.06(1+z)$ and identify 66,033
cluster candidates of $R_{L*}\ge20$ and $N_{200}\ge8$ at
$z\ge0.5$. The Stripe 82 region of SDSS covering 235 deg$^2$ has been
observed 2 mag deeper than other SDSS regions [19]. We
also identify clusters in the SDSS Stripe 82 region using the
photometric redshifts given in Ref. [20]. By setting the
photometric redshift slice of $z\pm0.04(1+z)$, 46,663 cluster/group
candidates have been identified with a lower threshold of $R_{L*}=10$
in the redshift of $0.05<z<1$.

We visually inspect images of above newly identified clusters to
search for lensed arcs. For clusters identified from SDSS DR8, we
inspect the composite $gri$ color
images\footnote{http://skyserver.sdss3.org/dr8/en/tools/chart/list.asp}.
For clusters in the Stripe 82 region, we inspect the black-white
images\footnote{http://cas.sdss.org/stripe82/en/tools/chart/list.asp}.
As done previously [10, 11], we search for arcs in images and consider
them as lensed background galaxies by a few criteria:\\
1. The arcs must be located in the center of a cluster, and must 
have a smoothly curved shape with respect to one or two central 
bright galaxies. Using this criterion, we can exclude cases for 
possible overlapping of a few faint galaxies. \\ 
2. The arcs have bluer colors than the cluster member galaxies. This
excludes possible faint member galaxies and ensures the arcs being
high-z star-forming galaxies. \\
3. The arcs have a good length-to-width ratio. A larger ratio
means a higher probability of true lensed background galaxies.\\ 
4. To exclude possible edge-on blue spiral galaxies, the arcs must be
faint and must not be associated with any bulges.
Blue spiral arms of some spiral galaxies can mimic arcs, but they are
usually extended from the central bulge and are symmetrically located
at two sides of the bulge. We have carefully check these feature and
exclude them from our sample.

Research efforts have been made by many international teams in the
development of algorithms for an automatic search of the lensed arcs,
involving a large amount of image processing and matching, for example
[21]. We prefer the simple manual inspections with the above
qualitative criteria, which is practically efficient and much less
resource-consuming.
In the images of about a hundred thousand galaxy clusters, 17 clusters
show giant arcs, including 152559.9+084639 and 162320.3+215535, which
were serendipitously discovered after publication of our catalog [11].
According to the goodness of arcs matching the above criteria, we
empirically classify these 17 systems into three classes: almost
certain lensing systems, probably systems, and possible cases.
There are 8 almost certain systems, 3 probably and 6 possible cases,
as listed in Table~\ref{tab1}. The arcs have a separation of
$\sim$2.5$''$ -- 14.8$''$ with respect to central cluster galaxies,
and most of them are bluer than the galaxies of clusters. We show the
images in Fig.~\ref{almost} for almost certain cases, in
Fig.~\ref{proba} and Fig.~\ref{possi} for probable and possible
cases, respectively.

\sec{3\quad Discussions}

\begin{figure}[H]
\centering
\includegraphics[width = 3.2in]{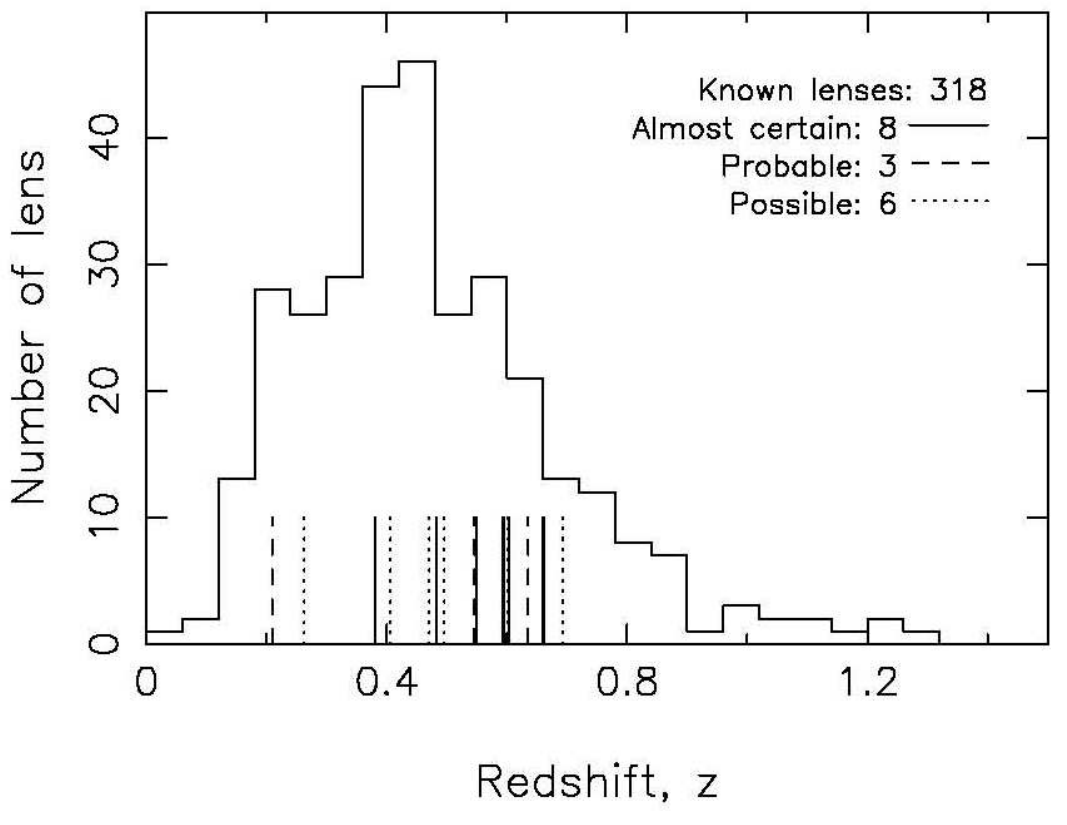}
\includegraphics[width = 3.2in]{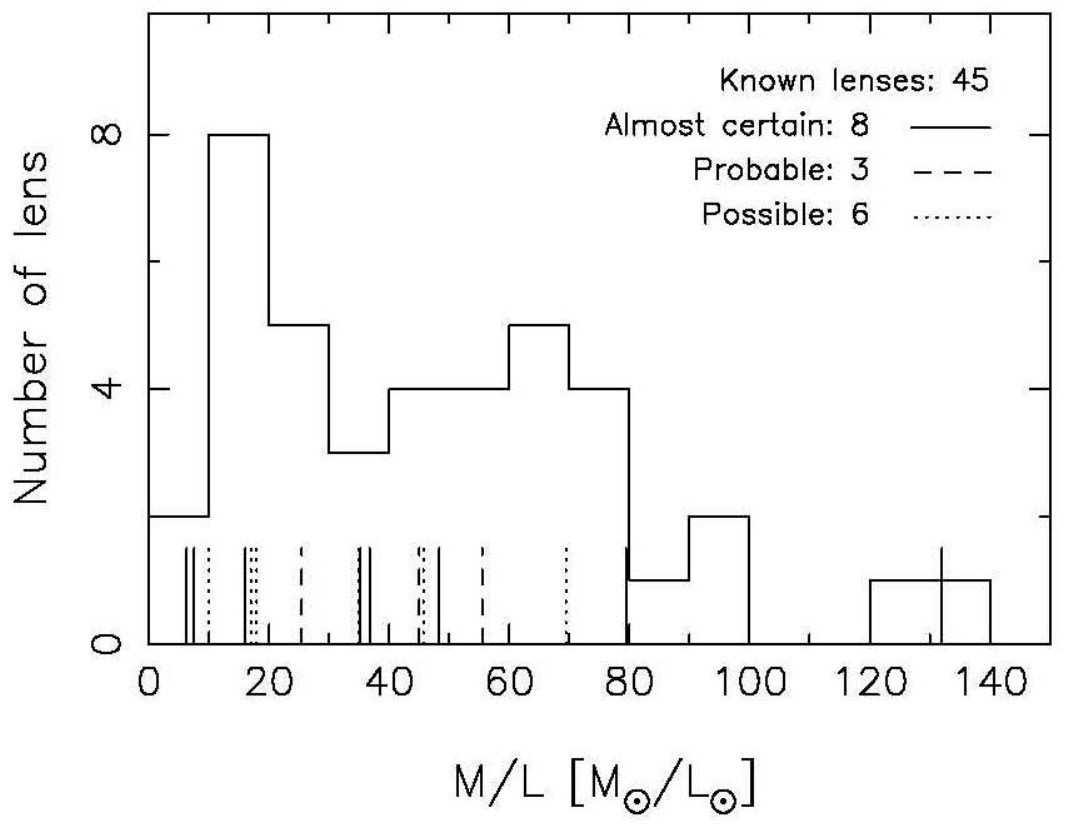}
\includegraphics[width = 3.2in]{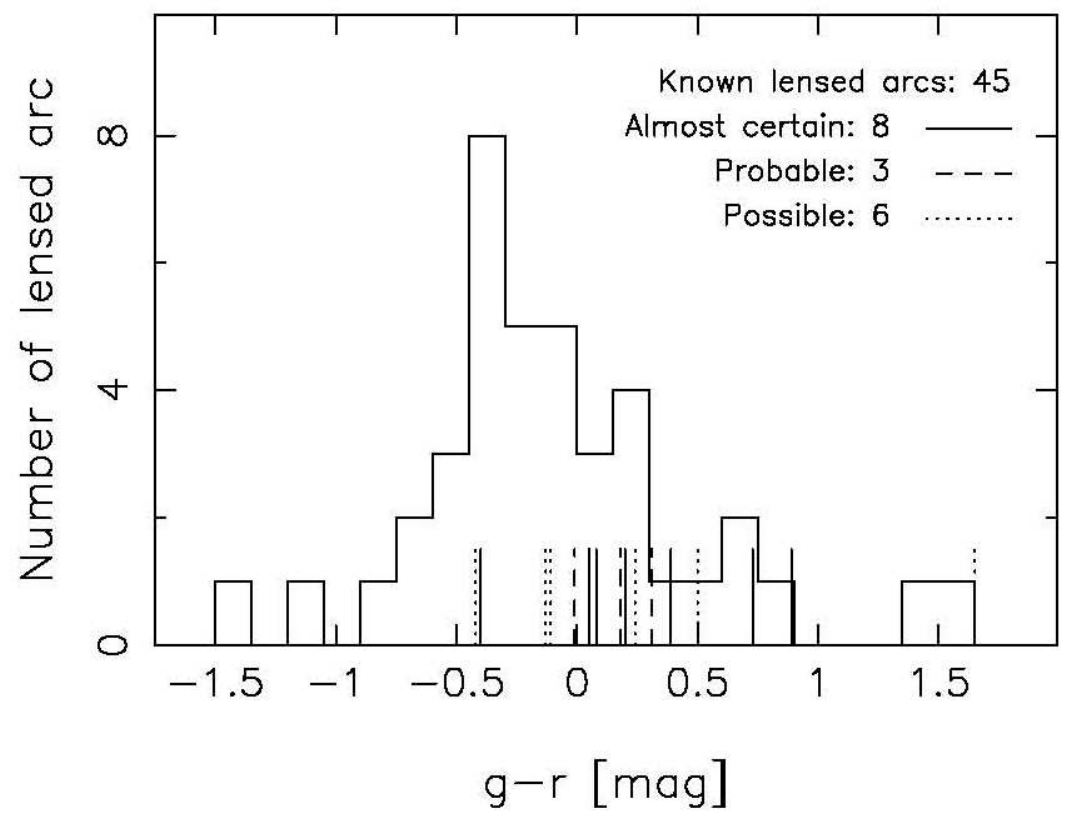}
\vspace{-2mm}
\caption{{\it Upper:} redshifts of our 17 lensing clusters are
  compared with the distribution of 318 known strong lensing
  clusters. {\it Middle and lower:} mass-to-light ratios and color
  ($g-r$) of our 17 lensing clusters of galaxies are compared with the
  values of 45 previously known lenses. }
\label{histogram}
\end{figure}

The newly found lensing clusters of galaxies on average have a higher
redshift than the 318 known strong lensing clusters, as shown in
Fig.~\ref{histogram}. The previously known lensing clusters are
collected from literature, such as [4, 6, 12--15, 22--24].

Based on the SDSS images and redshift data, we get the angular
separations between the arcs and the bright central galaxies. Here, we
assume the separation as being the Einstein radius, and then estimate
the mass within this radius. In a $\Lambda$CDM cosmology ($H_0=$70
${\rm km~s}^{-1}$ ${\rm Mpc}^{-1}$, $\Omega_m=0.275$ and
$\Omega_{\Lambda}=0.725$, hereafter), the mass is 
\begin{equation}
M(<\theta)=\frac{c^2\theta^2}{4G}\frac{D_{\rm l}D_{\rm s}}{D_{\rm ls}},
\label{lensm}
\end{equation}
where $D_{\rm s}$ and $D_{\rm l}$ are the angular diameter distances
of the source and lens from the observer, and $D_{\rm ls}$ is the
angular diameter distance of the source from the lens. We calculate
the total gravitational mass within the radius by assuming that the
source has a redshift $z_{\rm s}=1.5$. The values of the estimated
masses are listed in Table~\ref{tab1}.

We obtain the total $r$-band luminosity within the assumed Einstein
radius. We can calculate the mass-to-light ratios of the lensing
clusters, as listed in the last column of Table~\ref{tab1}. These
mass-to-light ratios are a few tens of the Solar value, which indicate
that there do exist a huge amount of invisible matter in galaxy
clusters, at least in the region inside the arcs. In
Fig.~\ref{histogram} we compared the mass-to-light ratios for our 17
lensing clusters with the values estimated from arcs of 45 previously
known lensing clusters in the SDSS DR8, and found that they are in the
same range and consistent.

The colors of the newly found arcs are also in the range of those of
arcs in 45 previously known lensing clusters (see the lower panel of
Fig.~\ref{histogram}).

Although the lensed arcs presented in this paper are found by visual
inspection of images of hundreds of thousands galaxy clusters, our
search of the arcs is efficient. The arcs we found do need
following-up observations to be confirmed as lensing systems, which
have to involve large optical telescopes for time-consuming
spectroscopic observations. We noticed that 12 cases (including 4
almost certain, 5 probable and 3 possible cases) of 13 lensing systems
in our first effort [10] have been confirmed by later spectroscopic
observations [13--15]. Recently, 16 case of our second effort [11]
have also been confirmed by [16]. We believe that the eight almost
certain cases in our third effort, maybe together with 3 probably and
6 possible cases, can be confirmed with further spectroscopic
observations.

\Acknowledgements{\bahao We thank anonymous referee for helpful
  comments. This work was supported by the National Natural Science
  Foundation of China (Grant Nos. 11103032 and 11261140641), the China
  Ministry of Science and Technology under grant No. 2013CB837900, and
  the Young Researcher Grant of National Astronomical Observatories,
  Chinese Academy of Sciences. }

\normalsize \vskip0.3in\parskip=0mm \baselineskip 18pt
\renewcommand{\baselinestretch}{1.1}\footnotesize\parindent=4mm\bahao

\REF{1\ }Allen S W, Evrard A E, Mantz A B. Cosmological parameters
from observations of galaxy clusters. Annual Review of Astronomy and
Astrophysics, 2011, 49:409--470

\REF{2\ }Wu X P, Fang L Z. A statistical comparison of cluster mass
estimates from optical/X-ray observations and gravitational lensing.
Astrophys J, 1997, 483:62--67

\REF{3\ }Wen Z L, Han J L, Liu F S. Mass function of rich galaxy
clusters and its constraint on $\sigma$$_{8}$. Mon Not Roy Astron Soc,
2010, 407:533--543

\REF{4\ }Luppino G A, Gioia I M, Hammer F, et al. A search for
gravitational lensing in 38 X-ray selected clusters of
galaxies. Astronomy and Astrophysics Supplement, 1999, 136:117--137

\REF{5\ }Gladders M D, Hoekstra H, Yee H K C, et al. The incidence of
strong-lensing clusters in the Red-Sequence Cluster Survey. Astrophys
J, 2003, 593:48--55

\REF{6\ }Sand D J, Treu T, Ellis R S, et al. A systematic search for
gravitationally lensed arcs in the Hubble Space Telescope WFPC2
Archive. Astrophys J, 2005, 627:32--52

\REF{7\ }Cabanac R A, Alard C, Dantel-Fort M, et al. The CFHTLS strong
lensing legacy survey. I. Survey overview and T0002 release
sample. Astronomy and Astrophysics, 2007, 461:813--821

\REF{8\ }Belokurov V, Evans N W, Moiseev A, et al. The cosmic
horseshoe: discovery of an einstein ring around a giant luminous red
galaxy. Astrophys J, 2007, 671:L9--L12

\REF{9\ }Belokurov V, Evans N W, Hewett P C, et al. Two new
large-separation gravitational lenses from SDSS. Monthly Notices of
the Royal Astronomical Society, 2009, 392:104--112

\REF{10\ }Wen Z L, Han J L, Xu X Y, et al. Discovery of four
gravitational lensing systems by clusters in the SDSS DR6. Research in
Astronomy and Astrophysics, 2009a, 9:5--10

\REF{11\ }Wen Z L, Han J L, Jiang Y Y. Lensing clusters of galaxies in
the SDSS-III. Research in Astronomy and Astrophysics, 2011,
11:1185--1198

\REF{12\ }Hennawi J F, Gladders M D, Oguri M, et al. A new survey for
giant arcs. Astronomical Journal, 2008, 135:664--681

\REF{13\ }Kubo J M, Allam S S, Annis J, et al. The Sloan Bright Arcs
Survey: six strongly lensed galaxies at z = 0.4-1.4. Astrophysical
Journal Letters, 2009, 696:L61--L65

\REF{14\ }Diehl H T, Allam S S, Annis J, et al. The Sloan Bright Arcs
Survey: four strongly lensed galaxies with redshift > 2. Astrophysical
Journal, 2009, 707:686--692

\REF{15\ }Bayliss M B, Wuyts E, Sharon K, et al. Two lensed lyman-α
emitting galaxies at z~5. Astrophysical Journal, 2010, 720:1559--1568

\REF{16\ }Stark D P, Auger M, Belokurov Vasily, et al. The CASSOWARY
spectroscopy survey: A new sample of gravitationally lensed galaxies
in SDSS. 2013, eprint arXiv:1302.2663

\REF{17\ }Wen Z L, Han J L, Liu F S. Galaxy clusters identified from
the SDSS DR6 and their properties. Astrophysical Journal Supplement,
2009b, 183:197--213

\REF{18\ }Wen Z L, Han J L, Liu F S. A catalog of 132,684 clusters of
galaxies identified from Sloan Digital Sky Survey III. Astrophysical
Journal Supplement, 2012, 199:34--46

\REF{19\ }Annis J, Soares-Santos M, Strauss M A, et al. The SDSS
Coadd: 275 deg$^2$ of deep SDSS imaging on Stripe 82. 2011, eprint
arXiv:1111.6619

\REF{20\ }Reis R R R, Soares-Santos M, Annis J, et al. The Sloan
Digital Sky Survey Co-add: a galaxy photometric redshift
catalog. Astrophysical Journal, 2012, 747:59--70

\REF{21\ }Estrada J, Annis J, Diehl H T, et al. A Systematic 
Search for High Surface Brightness Giant Arcs in a Sloan Digital 
Sky Survey Cluster Sample. Astrophysical Journal, 2007, 660:1176--1185

\REF{22\ }Bayliss M B, Hennawi J F, Gladders M D, et al. Gemini/GMOS
spectroscopy of 26 strong-lensing-selected galaxy cluster
cores. Astrophysical Journal Supplement, 2011, 193:8--34

\REF{23\ }More A, Cabanac R, More S, et al. The CFHTLS-Strong Lensing
Legacy Survey (SL2S): investigating the group-scale lenses with the
SARCS sample. Astrophysical Journal, 2012, 749:38--57

\REF{24\ }Furlanetto C, Santiago Bo X.,Makler,Ma, et . The SOAR
Gravitational Arc Survey -- I. Survey overview and photometric
catalogues. Monthly Notices of the Royal Astronomical Society, 2013,
432:73--88

\label{lastpage}
\end{multicols}

\end{document}